\documentclass[sigconf]{acmart}

\pdfoutput=1

\usepackage{rotating}
\usepackage{upquote}
\usepackage{amssymb}
\usepackage{graphicx}
\usepackage{comment}
\usepackage{listings}
\usepackage{subcaption}
\usepackage{multirow}
\usepackage{wasysym}
\usepackage{pifont}
\usepackage[draft]{todonotes} 
\usepackage{xcolor}
\usepackage{colortbl}
\usepackage{booktabs, cellspace, tabularx}
\usepackage{cramped}
\usepackage[T1]{fontenc}
\usepackage{enumitem}
\usepackage{epstopdf}
\setlist{leftmargin=5.5mm}

\copyrightyear{2019}
\acmYear{2019}
\acmConference[CCS '19]{2019 ACM SIGSAC Conference on Computer and Communications
Security}{November 11--15, 2019}{London, United Kingdom}
\acmBooktitle{2019 ACM SIGSAC Conference on Computer and Communications Security (CCS
'19), November 11--15, 2019, London, United Kingdom}
\acmPrice{15.00}
\acmDOI{10.1145/3319535.3363195}
\acmISBN{978-1-4503-6747-9/19/11}

\begin{document}
\fancyhead{}



\title[Security Certification in the Payment Card Industry]
{Security Certification in Payment Card Industry:\\ Testbeds, Measurements, and Recommendations}

\newcommand\Mark[1]{\textsuperscript#1}

\author{Sazzadur Rahaman$^1$,  Gang Wang$^2$, Danfeng (Daphne) Yao$^1$}
\affiliation{ 
     \institution{\Mark{1}Computer Science, Virginia Tech, Blacksburg, VA \\
     \Mark{2}Computer Science, University of Illinois at Urbana-Champaign, Urbana, IL}
    }
\email{sazzad14@vt.edu, gangw@illinois.edu, danfeng@vt.edu}

\newcommand{\pci}{{\sc PCI DSS}}
\newcommand{\cmark}{\ding{51}}%
\newcommand{\xmark}{\ding{55}}

\newcommand{\snote}[1]{\todo[inline,backgroundcolor=yellow!40]{\textbf{Sazzadur:} #1}} 
\newcommand{\dnote}[1]{\todo[inline,backgroundcolor=green!40]{\textbf{TODO:} #1}} 
\newcommand{\gnote}[1]{\todo[inline,backgroundcolor=blue!20]{\textbf{Gang:} #1}} 



\newcommand{\checker}{{\sc PciCheckerLite}}

\begin{abstract}

The massive payment card industry (PCI) involves various entities such as merchants, issuer banks, acquirer banks, and card brands. Ensuring security for all entities that process payment card information is a challenging task. The PCI Security Standards Council requires all entities to be compliant with the PCI Data Security Standard (DSS), which specifies a series of security requirements. However, little is known regarding how well PCI DSS is enforced in practice. In this paper, we take a measurement approach to systematically evaluate the PCI DSS certification process for e-commerce websites. We develop an e-commerce web application testbed, {\sc BuggyCart}, which can flexibly add or remove 35 PCI DSS related vulnerabilities. Then we use the testbed to examine the capability and limitations of PCI scanners and the rigor of the certification process. We find that there is an alarming gap between the security standard and its real-world enforcement. None of the 6 PCI scanners we tested are fully compliant with the PCI scanning guidelines, issuing certificates to merchants that still have major vulnerabilities. To further examine the compliance status of real-world e-commerce websites, we build a new lightweight scanning tool named \checker{} and scan 1,203 e-commerce websites across various business sectors. 
 The results confirm that 86\% of the websites have at least one PCI DSS violation that should have disqualified them as non-compliant. Our in-depth accuracy analysis also shows that \checker{}'s output is more precise than w3af. We reached out to the PCI Security Council to share our research results to improve the enforcement in practice.

\end{abstract}


\sloppy



\maketitle


{\fontsize{8pt}{8pt} \selectfont
\textbf{ACM Reference Format:}\\
Sazzadur Rahaman, Gang Wang, Danfeng (Daphne) Yao. 2019. Security Certification in Payment Card Industry: Testbeds, Measurements, and Recommendations. In \textit{2019 ACM SIGSAC Conference on Computer \& Communications Security (CCS'19), November 11--15, 2019, London, UK.} ACM, New York, NY, USA, 18 pages. https://doi.org/10.1145/3319535.3363195}

\section{Introduction}


Payment systems are critical targets that attract financially driven attacks. Major card brands (including Visa, MasterCard, American Express, Discover, and JCB) formed an alliance named Payment Card Industry Security Standards Council (PCI SSC) to standardize the security requirements of the ecosystem at a global scale.
The PCI Security Standards Council maintains, updates, and promotes Data Security Standard (DSS)~\cite{DBLP:books/pcidss/req} that defines a comprehensive set of security requirements for payment systems. PCI DSS certification has established itself as a global trademark for secure payment systems. According to PCI DSS~\cite{DBLP:books/pcidss/req},

\textit{``PCI DSS applies to \textbf{all} entities involved in payment card processing -- including merchants, processors, acquirers, issuers, and service providers. PCI DSS also applies to \textbf{all} other entities that store, process, or transmit cardholder data and/or sensitive authentication data.''}

The PCI Security Standards Council plays a major role in evaluating the security and compliance status of the payment card industry participants and supervises a set of entities that are responsible to perform compliance assessments such as Qualified security assessors (QSA) and Approved scanning vendors (ASV). All entities in the PCI ecosystem, including merchants, issuers, and acquirers, need to comply with the standards. PCI standards specify that entities need to obtain their compliance reports from the PCI authorized entities ({\em e.g.}, QSA and ASV) and periodically submit the reports in order to maintain their status. For example, a merchant needs to submit its compliance report to the acquirer bank to keep its business account active within the bank. Similarly, card issuer and acquirer banks need to submit their compliance reports to the payment brands ({\em e.g.}, Visa, MasterCard, American Express, Discover, and JCB) to maintain their membership status~\cite{DBLP:books/pcidss/req}.

However, several recent high-profile data breaches~\cite{DBLP:journals/corr/ShuTCY17, eqifaxbreach} have raised concerns about the security of the payment card ecosystem, specially for e-commerce merchants\footnote{Merchants that allow {\em online} payment card transactions for selling products and services are referred to as ``e-commerce merchants''.}.
A research report from Gemini Advisory~\cite{geminiy:creditcardfraud} shows that 60 million US payment cards have been compromised in 2018 alone. 
Among the merchants that experienced data breaches, many were known to be compliant with the PCI data security standards (PCI DSS). For example, in 2013, Target leaked 40 million payment card information due to insecure practices within its internal networks~\cite{DBLP:journals/corr/ShuTCY17}, despite that Target was marked as PCI DSS compliant. These incidents raise important questions about how PCI DSS is enforced in practice.  

In this paper, we ask: {\em how well are the PCI data security standards enforced in practice?} {\em Do real-world e-commerce websites live up to the PCI data security standards?} These questions have not been experimentally addressed before. We first design and develop testbeds and tools to quantitatively measure the degree of PCI DSS compliance of PCI scanners and e-commerce merchants. PCI scanners are commercial security services that perform external security scans on merchants' servers and issue certificates to those who pass the scan. By setting up our testbed, {\em i.e.}, an e-commerce website with configurable vulnerabilities, we empirically measure the capability of PCI scanners and the rigor of the certification process. 

Our results show that the detection capabilities of PCI scanners vary significantly, where even PCI-approved scanners fail to report serious vulnerabilities in the testbed. For 5 of the 6 scanners evaluated, the reports are not compliant with the PCI scanning guidelines~\cite{DBLP:books/asv/guideline}. All 6 scanners issued certificates to web servers that still have major vulnerabilities ({\em e.g.}, sending sensitive information over HTTP). Even if major vulnerabilities are detected ({\em e.g.}, remotely accessible MySQL), which should warrant an ``automatic failure'' according to the guideline~\cite{DBLP:books/asv/guideline}, some PCI scanners still proceed with certification regardless.

Given the weak scanner performance, it is possible that real-world e-commerce websites still have major vulnerabilities. For validation, we build a new lightweight scanning tool and perform empirical measurements on 1,203 real-world e-commerce websites.  
Note that for independent researchers or third-parties, scanning in the PCI context imposes a new technical challenge, namely the non-intrusive low-interaction constraint. The low interaction constraint, necessary for testing live production sites, makes it difficult to test certain vulnerabilities externally. Traditional penetration testing (pentest) tools are not suitable to test live websites in production environments. For example, pentest tools such as w3af~\cite{w3aflink} have brute-force based tests which require intense URL fuzzing ({\em e.g.}, prerequisite for SQL injection, XSS) or sending disruptive payload. The feedback from the PCI Security Council during our disclosure (Section~\ref{sec:discuss}) also confirmed this challenge. 




\noindent
Our technical contributions and findings are summarized below.


\begin{itemize}

    \item We design and develop an e-commerce web application testbed called {\sc BuggyCart}, where we implant 35 PCI-related vulnerabilities such as server misconfiguration ({\em e.g.}, SSL/TLS and HTTPS misconfigurations), programming errors ({\em e.g.}, CSRF, XSS, SQL Injection), and noncompliant practices ({\em e.g.}, storing plaintext passwords, PAN, and CVV). {\sc BuggyCart} allows us to flexibly configure vulnerabilities in the testbed for measuring the capabilities and limitations of PCI scanners. 
    
    \item Using {\sc BuggyCart}, we evaluated 6 PCI scanning services, ranging from more expensive scanners ({\em e.g.}, \$2,995/Year) to low-end scanners ({\em e.g.}, \$250/Year). The results showed an alarming gap between the specifications of the PCI data security standard and its real-world enforcement. For example, most of the scanners choose to certify websites with serious SSL/TLS and server misconfigurations. None of the PCI-approved scanning vendors detect SQL injection, XSS, and CSRF. 5 out of the 6 scanners are not compliant with the ASV scanning guidelines (Section~\ref{sec:findings}).
    
    
   \item We further evaluated 4 generic web scanners (not designed for PCI DSS), including two commercial scanners and two open-source academic solutions (w3af~\cite{w3aflink}, ZAP~\cite{zaplink}). We examine whether they can detect the web-application vulnerabilities missed by PCI scanners. Unfortunately, most of these vulnerabilities still remain undetected.  (Section~\ref{sec:findings}). 
    
    
    \item  We conducted empirical measurements to assess the (in)security of real-world e-commerce websites. We carefully designed and built a lightweight vulnerability scanner called \checker{}. 
    Our solution to addressing the non-intrusiveness challenge is centered at minimizing the number of requests that \checker{} issues per test case, while maximizing the test case coverage. It also involves a collection of lightweight heuristics that merge multiple security tests into one request.  Using \checker{}, we evaluated 1,203 e-commerce website across various business categories. We showed that 94\% of the websites have at least one PCI DSS violation, and 86\% of them contains violations that should have disqualified them as non-compliant (Section~\ref{sec:ourchecker}). 
    Our in-depth accuracy analysis also showed that \checker{}'s outputs have fewer false positives than the w3af counterpart (Table~\ref{result:pci-checker-vs-w3af}).
    
\end{itemize}

Based on our results, we further discuss how various PCI stakeholders, including the PCI council, scanning providers, banks, and merchants, as well as security researchers, can collectively improve the security of the payment card ecosystem (Section~\ref{sec:discuss}). We open-sourced our {\sc BuggyCart}~\footnote{Available at https://github.com/sazzad114/buggycart} testbed and {\sc PciCheckerLite}~\footnote{Available at https://github.com/sazzad114/pci-checker} in GitHub, which also include a pre-installed docker image. We are in the process of sharing {\sc BuggyCart} with the PCI security council (Section~\ref{sec:method}).

\section{Background on PCI and DSS}
\label{sec:back}
We start by describing the background for the security practices, workflow, and standards of the current PCI ecosystem that involves banks, store-front and e-commerce vendors, and software providers.  Then, we focus on how merchants obtain security certifications and establish trust with the banks. We discuss how the certification process is regulated and executed.  


\begin{figure}[t]
   \centering
 	\includegraphics[width=0.47\textwidth]{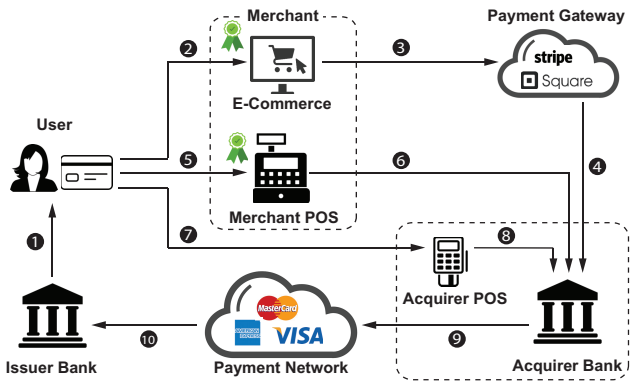}
  	\caption{Overview of the payment card ecosystem. }\label{fig:payment-card-ecosystem}
  	\vspace{-0.18in}
 \end{figure}
 

 \subsection{Payment Card Ecosystem}
 The Payment Card Industry (PCI) has established a working system that allows merchants to accept user payment via payment cards, and complete the transactions with the banks in the backend. Figure~\ref{fig:payment-card-ecosystem} shows the relationships between the key players in the ecosystem, including users, merchants, and banks. The user and the merchant may use different banks. The {\em issuer bank} issues payment cards to the user and manages the user's credit or debit card accounts (step \ding{182}). Users use the payment card at various types of merchants (steps \ding{183}, \ding{186}, and \ding{188}). The \textit{acquirer bank} manages an account for the merchant to receive and route  the transaction information (steps \ding{185}, \ding{187}, and \ding{189}). The acquirer bank ensures that funds are deposited into the merchant's account once the transaction is complete via the payment network (steps \ding{190} and \ding{191}). The payment network, also known as the card brands ({\em e.g.}, Visa, MasterCard), bridges between the acquirer and the issuer banks. 
 
There are different types of merchants. For merchants that run an e-commerce service ({\em i.e.}, all transactions are made online),  they usually interact with the acquirer bank via a \textit{payment gateway} ({\em e.g.}, Stripe, Square), which eases the payment processing and integration (\ding{184}). For merchants that have a physical storefront, they use point-of-sale (POS) devices, {\em i.e.}, payment terminals, to collect and transfer user card information to the acquirer bank. They can use either the acquirer bank's POS (\ding{188}) or their own POS (\ding{186}). The key difference is that acquirer POS directly transfers the card information to the bank without storing the information within the merchant. Merchant POS, however, may store the card information. 

Due to the fact that e-commerce websites and merchant POSes need to store card information, the merchants need to prove to the bank that they are qualified to securely handle the information processing. The acquirer bank requires these merchants to obtain PCI security certifications in order to maintain accounts with the bank~\cite{pciquestionnaire}. Next, we introduce the security certification process.

\subsection{PCI Council and Data Security Standard}
\label{sec:dss-back}
Payment Card Industry Security Standards Council manages a number of specifications to ensure data security across the extremely complex payment ecosystem. Among all the specifications, only the Data Security Standard (DSS) and Card Production and Provisioning (CPP) are {\em required}. All the other specifications (shown in Table~\ref{pcissc:specs} in the Appendix) are recommended ({\em i.e.}, optional). CPP is designed to regulate card issuers and manufactures. The Data Security Standard (DSS) is the most important specification that is required to be complied by issuer banks, acquirer banks, and all types of merchants and e-commerce sites, {\em i.e.},  all systems that process payment cards. Our work is focused on the DSS compliance. 

In the PCI Data Security Standard specifications~\cite{DBLP:books/pcidss/req}, there are 12 requirements that an organization must follow to protect user payment card data. These requirements cover various aspects ranging from network security to data protection policies, vulnerability management, access control, testing, and personnel management. In total, there are 79 more detailed items under the 12 high-level requirements. We summarize them in the Appendix (Table~\ref{test-cases}). 

DSS applies to all players in the ecosystem, including {\em all merchants and acquirer/issuer banks}. For merchants, they need to approve their compliance to the acquirer bank to open an account for their business. For acquirer and issuer banks, they need to prove their compliance to the card brands ({\em e.g.,} Visa, MasterCard) for the eligibility of membership.  


We use the merchant as an example to illustrate how DSS compliance is assessed. 
First, the PCI security standard council provides the specifications and self-assessment questionnaires (SAQ)~\cite{pciquestionnaire}. Merchants self-assess their DSS compliance and attach the questionnaires in their reports. Second, the merchant must pass the security tests and audits from external entities such as Approved Scanning Vendors (ASV) and the Qualified Security Assessors (QSA). The PCI council approves a list of ASV and QSA~\cite{pciasv} for the assessment. 

 \begin{table}[t]
\caption{PCI Compliance levels and their evaluation criteria.}~\label{compliance-table}
\vspace{-0.1in}
\centering
\small
\begin{tabular}{lllll}
\toprule
\multirow{2}{*}{Level} & \multirow{2}{*}{Transactions} & \multicolumn{3}{c}{Compliance Requirements} \\ \cline{3-5} 
& Per Year & \begin{tabular}[c]{@{}l@{}} Self-report \\ with SAQ \end{tabular} &
\begin{tabular}[c]{@{}l@{}}Sec Scans  \\ by ASV \end{tabular} &
\begin{tabular}[c]{@{}l@{}}Sec Audits \\ by QSA\end{tabular}  \\ \cline{1-5}
Level 1 & Over 6M & Quarterly & Quarterly & Required  \\
Level 2 & 1M -- 6M & Quarterly & Quarterly & Required/Optional  \\
Level 3 & 20K -- 1M & Quarterly & Quarterly & \textbf{Not Required} \\
Level 4 & Less than 20K & Quarterly & Quarterly & \textbf{Not Required} \\ 
\bottomrule
\end{tabular}
\vspace{-0.15in}
\end{table}

\noindent
\textbf{Security scanning} is conducted by certified scanners (Approved Scanning Vendors or ASVs) on card processing entities. Security scanning is performed remotely without the need for on-site auditing. Not all the requirements can be automatically verified by the remote scanning (see Table~\ref{pcissc:specs} in the Appendix). The PCI council provides an ASV scanning guideline~\cite{DBLP:books/asv/guideline}, which details the responsibilities of the scanners (see Table~\ref{asv:scanning:guideline} in the Appendix).
 
\noindent
\textbf{Self-assessment questionnaires (SAQs)} allow an organization to self-evaluate its security compliance~\cite{DBLP:books/saq/guideline}. In SAQs, all the questions are close ended. More SAQ analysis is presented in Section~\ref{sec:discuss}. 

\noindent
\textbf{Security audit} is carried out by Qualified Security Assessors (QSAs). It requires on-site auditing ({\em e.g.}, checking network and database configurations, examining software patches, and interviewing employees). As security scanning cannot verify all of the DSS properties, on-site audits are to cover those missing aspects.

\noindent
\textbf{Level of compliance} varies for different organizations. The compliance level is usually determined by the number of annual financial transactions handled by the organization. Each acquirer bank (or card brand) has its own program for compliance and validation levels. In Table~\ref{compliance-table}, we show the tentative compliance levels that roughly match most of the payment brands~\cite{pcimerchantlevel, pciquestionnaire}. The self-assessment questionnaires (SAQs) and security scanning are required quarterly regardless of the compliance levels. Only large organizations that handle over 1 million transactions a year are required to have the on-site audit (by a QSA). The majority of merchants are small businesses, ({\em e.g.}, 85\% of merchants all over the world have less than 1 million USD web sale~\cite{ecommercemarketsize}). Thus, most online merchants rely on ASV scanners and self-reported questionnaires for compliance assessment.

\subsection{Our Threat Model and Method Overview}

\noindent
\textbf{Threat Model.} 
%
The certification process is designed as an enforcement mechanism for merchants to hold a high-security standard to protect user data from external adversaries. If the certification process is not well executed, it would allow merchants with security vulnerabilities to store payment card data and interact with banks. In addition, such security certification may also create a false sense of security for merchants. We primarily focus on the automatic server screening by PCI scanners given that all merchants need to pass the scanning. We also briefly analyze the Self-assessment Questionnaire (SAQs). Our analysis does not cover on-site audits, because {\em i)} on-site audit is not required for the vast majority of the merchants and {\em ii)} it is impossible to conduct analysis experiments on on-site audits without partnerships with service providers.

\noindent
\textbf{Methodology Overview.} To systematically measure and compare the rigor of the compliance assessment process, our methodology is to build a semi-functional e-commerce website as a testbed and order commercial PCI scanning services to screen and certify the website. The testbed allows us to easily configure website instances by adding or removing key security vulnerabilities that PCI DSS specifies. We leverage this testbed to perform controlled measurements on the certification process of a number of PCI scanners. In addition to the controlled experiments, we also empirically measure the security compliance of real-world e-commerce websites with a focus on a selected set of DSS requirements. In the following, we describe our detailed measurement methodologies and findings.

\section{Measurement Methodology}
\label{sec:method}
In this section, we describe our measurement methodology for understanding how PCI scanners perform data security standard (DSS) compliance assessment and issue certificates to merchants. The core idea is to build a re-configurable testbed where we can add or remove key security vulnerabilities related to DSS and generate testing cases. By ordering PCI scanning services to scan the testbed, we collect incoming network traffic as well as the security compliance reports from the scanning vendors. In the following, we first describe the list of vulnerabilities that our testbed covers, and then introduce the key steps to set up the e-commerce frontend. 

\subsection{Security Test Cases}
\label{experimental-design}
The testbed contains a total of 35 test cases, where each test case represents a type of security vulnerabilities. Running a PCI scanner to scan the testbed could reveal vulnerabilities that the scanner can detect, as well as those that the scanner fails to report. We categorize the 35 security test cases $i_1$--$i_{35}$ into four categories, namely {\em network security}, {\em system security}, {\em application security}, and {\em secure storage}. Note that there are 29 test cases in the first three categories are within the scope of ASV scanners ({\em i.e.}, ASV testable cases). The other 6 cases under ``secure storage'' cannot be remotely verified. We include these cases to illustrate the limits of ASV scanners.  

\begin{enumerate}
    \item {\bf Network security (14 test cases).} These testing cases are related to network security properties, including firewall status,  ($i_{1}$), the access to critical software from network ($i_{2}$--$i_{4}$), default passwords ($i_{5}$--$i_{6}$), the usage of HTTP to transmit sensitive data ({\em e.g.}, customer or admin login information) ($i_{7}$), and SSL/TLS misconfigurations ($i_{12}$--$i_{18}$).
    
    \item {\bf System security (7 test cases).} These test cases are related to system vulnerabilities, including vulnerable software ($i_{19}$--$i_{20}$), server misconfigurations ($i_{29}$--$i_{32}$), and HTTP security headers ($i_{33}$). 
    
    \item {\bf Web Application security (8 test cases).} These test cases are related to application-level problems including SQL injections ($i_{21}$--$i_{22}$), not following secure password guidelines ($i_{23}$--$i_{24}$), the integrity of Javascripts  from external sources ($i_{25}$), revealing crash reports ($i_{26}$), XSS ($i_{27}$) and CSRF ($i_{28}$).
    
    \item {\bf Secure storage (6 test cases).} Secure storage is impossible to verify through external scans. Thus, DSS does not require PCI scanners to test these properties, such as storing sensitive user information ($i_{8}$), storing and showing PAN in plaintext ($i_{9}$--$i_{11}$), and insecure ways of storing passwords ($i_{34}$--$i_{35}$). In PCI DSS, merchants need to fill out the self-assessment questionnaire about how they handle sensitive data internally. We choose to include these vulnerabilities in the testbed for highlighting the fundamental limitations of external scans on some important aspects of server security. 
\end{enumerate}




\noindent
\textbf{Must-fix Vulnerabilities.} These test cases are designed following the official ASV scanning guideline~\cite{DBLP:books/asv/guideline} and the PCI data security standard (DSS)~\cite{DBLP:books/pcidss/req}. Among the 35 cases, 29 are within the scope (responsibility) of ASV scanners that can be remotely tested. After vulnerabilities are detected, website owners are required to fix any vulnerabilities that have a CVSS score $\geq$ 4.0, and any vulnerabilities that are marked as mandatory in PCI DSS. CVSS (Common Vulnerability Scoring System) measures the severity of a vulnerability (score 0 to 10). The CVSS scores in Table~\ref{table:experimental-results} are calculated using CVSSv3.0 calculator~\cite{cvss-v3-calculator}. Vulnerabilities that have no CVSS score are marked as ``N/A''. If the website owner fails to resolve the ``must-fix'' vulnerabilities, a scanner should not issue the compliance certification. As shown in Table~\ref{table:experimental-results}, 26 out of the 29 testable cases are required to be fixed. Three cases (vulnerability 3, 4, and 18) are not mandatory to fix. For example, exposing OpenSSH to the Internet (case-3) does not mean immediate danger as long as the access is well protected by strong passwords or SSH keys. 

\noindent
\textbf{Completeness and Excluded Cases.} 
When building our BuggyCart testbed and the \checker{} prototype, we exclude five mandatory ASV scanning cases: {\em i)} backdoors or malware,
{\em ii)} DNS server vulnerabilities, {\em e.g.}, unrestricted DNS zone transfer, 
{\em iii)} vulnerabilities in mail servers, 
{\em iv)} vulnerabilities in hypervisor and virtualization components, and  {\em v)} vulnerabilities in wireless access points. Most of them (namely, {\em ii}, {\em iii}, {\em iv}, and {\em v}) are not relevant, as they involve servers or devices outside our testbed or an application server. In the first category, it is difficult to design a generic network-based testing case. We also exclude the non-mandatory cases (shown in the last 4 rows of Table~\ref{asv:scanning:guideline} in the Appendix). 

Note that ASV testable cases only represent a subset of PCI DSS specifications~\cite{DBLP:books/pcidss/req} because some specifications are not remotely verifiable. There are specifications related to organization policies, which are impossible to verify externally, {\em e.g.}, ``restricting physical access to cardholder data'' (DSS req. 9), and ``documenting the key management process'' (DSS req. 3.6). They can only be assessed by onsite audits, which unfortunately are not applicable to the majority of e-commerce websites and small businesses (see Table~\ref{compliance-table}). We will discuss this further in Section~\ref{sec:discuss}.

Our \checker{} prototype in Section~\ref{sec:ourchecker} scans 17 test cases in Table~\ref{table:checkerresult}, which are a subset of the 29 externally scannable rules in Table~\ref{table:experimental-results}. When scanning live production websites, we have to eliminate cases that require intrusive operations
such as web crawling, URL fuzzing, or port scanning.






\subsection{Testbed Architecture and Implementations}

A key challenge of measuring PCI scanners is to interact with PCI scanners like a real e-commerce website does, in order to obtain reliable results. This requires the testbed to incorporate most (if not all) of the e-commerce functionality to interact with PCI scanners and reflect the scanners' true performance. For this reason, we choose the OpenCart~\cite{OpenCart} as the base to build our testbed. OpenCart is a popular open source PHP-based e-commerce solution used by real-world merchants to build their websites. This allows us to interact with PCI scanners in a realistic manner to ensure the validity of measurement results.   

\noindent
\textbf{Testbed Frontend.} The frontend of our testbed supports core e-commerce functionality, such as account registration, shopping cart management, and checkout and making payment with credit cards. The code of the website\footnote{The URL was {\tt www.rwycart.com}. We took the site offline after the experiment.} is based on {\sc OpenCart}. We rewrote the OpenCart system by integrating all 35 security vulnerabilities and testing cases. We deployed the website using Apache HTTP server and MySQL database.  Our testbed automatically spawns a website instance following a pre-defined configuration. We used OpenSSH as the remote access software and Phpmyadmin to remotely manage the MySQL database. We hosted our website in Amazon AWS in a single \textit{t2.medium} server instance with Ubuntu 16.04. We obtained a valid SSL certificate to enable HTTPS from {\em Let's Encrypt}~\cite{letsencrypt}.

We set up the website solely for research experiment purposes. Thus, it does not have a real payment gateway. Instead, we set up a dummy payment gateway that imitates the real gateway Cardconnect~\cite{cardconnect}. The website forwarded credit card transactions to this dummy payment gateway. The dummy endpoint for Cardconnect is implemented using \textit{flask-restful} framework. We modified the \texttt{/etc/hosts} file of our web server to redirect the request. During our experiments, our server did not receive any real payment transaction requests. We further discuss research ethics in Section~\ref{sec:ethic}.

\noindent
\textbf{Implementing Security Test Cases.} Next, we describe the implementation details of the 35 security test cases in Table~\ref{table:experimental-results}.

For the network security category, we implement test cases $i_1$ to $i_3$ by changing inbound traffic configurations within the Amazon AWS security group. Test case $i_4$ (administer access over Internet) is implemented by changing \textit{phpmyadmin} configuration. For test case $i_5$ (default SQL password), we do not set any password for ``\textit{root}'' and enable access from {\em any remote host}. Test case $i_5$ is implemented by configuring phpmyadmin (no password for user ``\textit{root}''). Test case $i_7$ is set to keep port 80 (HTTP) open without a redirection to port 443 (HTTPS). Test cases $i_{12}$, $i_{14}$, $i_{16}$, and $i_{17}$ are implemented by using default certificates from Apache. Test cases $i_{13}$ and $i_{18}$ are implemented by changing \textit{SSLCipherSuite} and \textit{SSLProtocol} of the Apache server. For test case $i_{15}$, we configure the Apache server to use a valid certificate but with a wrong domain name.

For the system security category, we implement test cases $i_{19}$--$i_{20}$ by installing software that are known to be vulnerable. For test case $i_{19}$, we use \textit{OpenSSL 7.2}, which is vulnerable to privilege escalation and timing side channel attacks. For test case $i_{20}$, we used \textit{phpmyadmin 4.8.2} which is known to be vulnerable to XSS. We implemented test cases $i_{29}$ to $i_{33}$ by changing the configurations of the Apache server. For test case $i_{33}$ (HTTP security header) in particular, we consider X-Frame-Options, X-XSS-Protection, X-Content-Type-Options, and Strict-Transport-Security.

For the web  application security category, we implement test cases $i_{21}$ to $i_{28}$ by modifying OpenCart source code~\cite{OpenCart}. Regarding secure password guidelines, we disable password retry restrictions for both users and administrators (test case $i_{23}$), disable the length checking of passwords (test case $i_{24}$). For SQL injection, we modify the admin login (test case $i_{21}$) and customer login (test case $i_{22}$) code to implement SQL injection vulnerabilities. For admin login, we simply concatenate user inputs without sanitation for the login query. For the customer login, we leave an SQL injection vulnerability at the login form. Given that the user password is stored as unsalted MD5 hashes, we run the login query by concatenating the MD5 hash of the user-provided password, which is known to be vulnerable to SQL injection~\cite{rawmd5sqlinjection}. 
For XSS and CSRF, we implant an XSS vulnerability in the page of editing customer profiles, by allowing HTML content in the ``first name'' field (test case $i_{27}$). By default, Opencart does not have any protection against CSRF (test case $i_{28}$). For test case $i_{26}$ (displaying errors), we configure OpenCart to reveal crash reports (an insecure practice, which gives away sensitive information). Opencart by default does not check the integrity of Javascript code loaded from external sources (test case $i_{25}$). 

For the secure storage category, we modify the Cardconnect extension to store CVV in our database (test case $i_8$) and the full PAN (instead of the last 4 digits) in the database in plaintext (test case $i_{10}$). We add an option to encrypt PANs before storing, but the encryption key is hardcoded (test case $i_{11}$). We also update the customers' order history page to show the unmasked PAN for each transaction (test case $i_{9}$). Finally, the testbed stores the raw unsalted MD5 hash of passwords for customers (test case $i_{34}$) and plaintext passwords for admins (test case $i_{35}$).

\subsection{Research Ethics}
\label{sec:ethic}
We have taken active steps to ensure research ethics for our measurement on PCI scanners (Section~\ref{sec:findings}). Given that our testbed is hosted on the public Internet, we aim to prevent real users from accidentally visiting the website (or even putting down credit card information). First, we only put the website online shortly before the scanning experiment. After each scanning, we immediately take down the website. Second, the website domain name is freshly registered. We never advertise the website (other than giving the address to the scanners). 
Third, we closely monitor the HTTP log of the server. Any requests ({\em e.g.}, for account registration or payment) that are not originated from the scanners are dropped. Network traffic from PCI scanners are easy to distinguish (based on IP and User-Agent) from real user visits. We did not observe any real user requests or payment transactions during our experiments.

All PCI scanners run automatically without any human involvement from the companies. We order and use the scanning services just like regular customers. We never actively generate traffic to the scanning service, and thus our experiments do not cause any interruptions. Our experiments follow the terms and conditions specified by the scanning vendors, which we carefully examined. We choose to anonymize the PCI scanners' names since some scanning vendors strictly forbid publishing any benchmark results. We argue that publishing our work with anonymized scanner names is sufficient for the purpose of describing the current security practice in the payment card industry, as the security issues reported are likely industry-wide, not unique to the individual scanners evaluated. In addition, anonymization would help alleviate the bias toward individual scanners and potential legal issues~\cite{DBLP:conf/ccs/Gamero-GarridoS17}.


In Section~\ref{sec:ourchecker}, we also carefully design our experiments when evaluating the compliance of 1,203 websites. The experiment is designed in a way that generates minimal footprints and impact on the servers, in terms of the number of connection requests to the servers. Our client is comparable to a normal client and does not cause any disruption to the servers. For example, we quickly closed the connection, after finding out whether or not an important port is open. More details are be presented in Section~\ref{sec:ourchecker}. 



\begin{figure}[t]
   \centering
 	\includegraphics[width=0.38\textwidth]{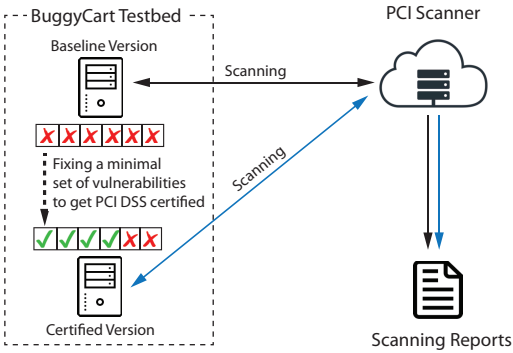}
  	\caption{Illustration of the baseline scanning and the certified version. A PCI scanner iteratively scans the testbed. The initial scan (baseline) is on the original testbed with all 35 vulnerabilities. The certified version is the testbed version where the testbed successfully passes the scanning after we iteratively fix a minimal set of vulnerabilities in the testbed. In Table~\ref{table:experimental-results}, we report the scanning results on both versions of the testbed for each scanner.}\label{fig:exp}
  	\vspace{-0.2in}
\end{figure}

\section{Evaluation of PCI Scanners}
\label{sec:findings}

Our first set of experiments is focused on evaluating PCI scanners to answer the following research questions. Later in Section~\ref{sec:ourchecker}, we will introduce our second set of experiments on measuring the security compliance of real-world e-commerce websites.

\begin{itemize}
    \item How do various PCI scanners compare in terms of their detection capabilities?  (Section~\ref{sec:asv_eval})
    \item What are the security consequences of inadequate scanning and premature certification? (Section~\ref{sec:securityissues}) 
    \item How are web scanners (commercial or open-source ones) compared with PCI scanners in terms of detection capabilities? (Section~\ref{sec:eval-website-scanners})
\end{itemize}

We selected 8 U.S. based PCI DSS scanners as shown in Table~\ref{tab:spent}. 
The selection process is as follows. From the list of approved vendors~\cite{pciasv}\footnote{As of April 30, 2019, 97 companies are approved by the PCI Council as the approved scanning vendors (ASVs)~\cite{pciasv}.}, we found 85 of them operate globally. Out of these 85, we aimed to identify a set of ASVs that appear to be of high quality ({\em e.g.}, judging from the company's reputations and websites) and somewhat affordable (due to our limited funding), while also covering different price ranges. We identified 6 such scanners. For 3 of them, the prices are publicly available. For the other 3 scanners, we emailed them through our \url{rwycart.com} email addresses. 2 of them (Scanner7 and Scanner8) did not provide their price quotations, which forced us to drop them from our evaluation (due to our organization policies). During our search, we also found that some website owners used non-ASV scanners. Thus, we also included 2 non-ASVs that have good self-reported quality. Non-approved scanners offer commercial PCI scanning services, but are not on the ASV list~\cite{pciasv} of the PCI council\footnote{To become an ASV, a scanner service needs to pay a fee and go through a testbed-based approval evaluation supervised by the PCI Council.}. Because of the legal constraints imposed by the terms and conditions of scanners, we cannot reveal scanners' names. Researchers who wish to reproduce or extend our work for scientific purposes without publishing scanner names are welcome to contact the authors.

%

We conducted experiments successfully with 6 of the scanners (without Scanner7 and Scanner8 for the reason mentioned above). We use the email address ({\tt wayne@rwycart.com}) associated with the testbed e-commerce website to register accounts at the scanning vendors. Table~\ref{tab:spent} shows the prices of these 6 vendors. For Scanner2 and Scanner1, we completed our experiments within the trial period (60 days for Scanner2 and 30 days for Scanner1). The trial-version and the paid-version offer the same features and services.

\begin{table}[t]
\caption{Prices of PCI scanners and the actual costs.}\label{tab:spent}
\vspace{-0.1in}
\small
\begin{tabular}{@{}llll@{}}
\toprule
PCI Scanners &  Price        & Spent Amount & PCI SSC Approved?\\ \midrule
Scanner1      & \$2,995/Year        & \$0 (Trial)     & Yes    \\
Scanner2     &  \$2,190/Year        & \$0 (Trial)    & Yes      \\
Scanner3  & \$67/Month   & \$335  & No      \\
Scanner4    & \$495/Year   & \$495  & Yes      \\
Scanner5      & \$250/Year   & \$250  & Yes      \\
Scanner6  & \$59/Quarter & \$118  & No      \\
\midrule
Scanner7         & Unknown       & N/A  & Yes     \\
Scanner8 &   \$350/Year  & N/A   & Yes    \\
\midrule{}
Total  & -   & \$1198   & -     \\
\bottomrule
\end{tabular}
\vspace{-0.15in}
\end{table}



\definecolor{Gray}{gray}{0.85}
\newcolumntype{a}{>{\columncolor{Gray}}c}
\newcolumntype{?}{!{\vrule width 1.5pt}}
\begin{table*}[th!]

\renewcommand{\aboverulesep}{0pt}
\renewcommand{\belowrulesep}{0pt}
\renewcommand{\arraystretch}{1.2}
\setlength{\tabcolsep}{0.4em}

\caption{Testbed scanning results. ``Baseline'' indicates the scanning results on our testbed when all the 35 vulnerabilities are active. ``Certified'' indicates the scanning results after fixing the minimum number of vulnerabilities in order to be compliant. ``\Circle'', ``\LEFTcircle'', ``\CIRCLE'' means severity level of low, medium, and high respectively according to the scanners. ``\xmark'' mean ``undetected'', ``\cmark'' means ``fixed in the compliant version'', ``\cmark$^*$'' means ``fixed as a side-effect of another case''. The ``website scanners'' represent a separate experiment to determine whether website scanners can help to improve coverage. We ran the website scanners on test cases that were not detected by the PCI ASV scanners. ``N/A'' means "not testable by an external scanner". ``-'' means "testable but do not need to tested". The "Must Fix" column shows the vulnerabilities that must be fixed by the e-commerce websites in order to be certified as PCI DSS compliant. 
}\label{table:experimental-results}
\vspace{-0.1in}
\centering
\footnotesize
\begin{tabular}{@{}llll|l|l?>{\centering\columncolor{Gray}}p{0.44cm}>{\centering}p{0.44cm}|>{\centering\columncolor{Gray}}p{0.44cm}>{\centering}p{0.44cm}|>{\centering\columncolor{Gray}}p{0.44cm}>{\centering}p{0.44cm}|>{\centering\columncolor{Gray}}p{0.44cm}>{\centering}p{0.44cm}|>{\centering\columncolor{Gray}}p{0.44cm}|c?>{\centering}p{0.44cm}|>{\centering}p{0.44cm}|l|l@{}}
\toprule
\multirow{2}{*}{Rq.} & \multirow{2}{*}{Test Cases}                                & \multirow{2}{*}{\begin{tabular}[c]{@{}l@{}}Vul.\\ Location\end{tabular}} & 
\multirow{2}{*}{\rotatebox[origin=c]{90}{\begin{tabular}[c]{@{}l@{}}~In ASV Scope?\end{tabular}}} &
\multirow{2}{*}{\rotatebox[origin=c]{90}{\begin{tabular}[c]{@{}l@{}}CVSS Score\end{tabular}}} &
\multirow{2}{*}{\rotatebox[origin=c]{90}{\begin{tabular}[c]{@{}l@{}}Must Fix?\end{tabular}}} &
\multicolumn{2}{c|}{Scanner2} & \multicolumn{2}{c|}{Scanner5} & \multicolumn{2}{c|}{\begin{tabular}[c]{@{}l@{}}Scanner4\\ / Scanner1\end{tabular}} & \multicolumn{2}{c|}{\begin{tabular}[c]{@{}l@{}}Scanner6\\ (not aprvd.) \end{tabular}} & \multicolumn{2}{c?}{\begin{tabular}[c]{@{}l@{}}Scanner3\\ (not aprvd.) \end{tabular}} &
\multicolumn{4}{c}{\begin{tabular}[c]{@{}l@{}}Website\\ Scanners \end{tabular}}\\ 
\cmidrule{7-20}
                 & & & & & & \rotatebox[origin=c]{90}{~Baseline~}             & \rotatebox[origin=c]{90}{~Certified~}            & \rotatebox[origin=c]{90}{~Baseline~}             & \rotatebox[origin=c]{90}{~Certified~}            & \rotatebox[origin=c]{90}{~Baseline~}              & \rotatebox[origin=c]{90}{~Certified~}             & \rotatebox[origin=c]{90}{~Baseline~}               & \rotatebox[origin=c]{90}{~Certified~}              & \rotatebox[origin=c]{90}{~Baseline~}               & \rotatebox[origin=c]{90}{~Certified~} & \rotatebox[origin=c]{90}{~Scanner2W~}
                 & \rotatebox[origin=c]{90}{~Scanner5W~} & \rotatebox[origin=c]{90}{~W3af~}  & \rotatebox[origin=c]{90}{~ZAP~}
                 \\ \midrule
    1.1 & 1. Firewall detection & OS & Y & N/A & Y  & \Circle             & \Circle            & \Circle            & \Circle            & \Circle             & \Circle             & \xmark              & \xmark             & \xmark              & \xmark & - & - & - & - \\  \midrule

\multirow{3}{*}{1.2} & 2. Mysql port (3306) detection & OS  & Y & N/A & Y & \CIRCLE & \cmark & \LEFTcircle & \cmark & \Circle & \Circle & \Circle  & \Circle & \Circle & \Circle & - & - & - & -  \\ \cmidrule(l){2-20}

 & 3. OpenSSH detected  & OS & Y & N/A & N & \LEFTcircle  & \cmark            & \Circle            & \Circle            & \Circle             & \Circle             & \Circle              & \Circle              & \Circle              & \Circle & - & - & - & - \\ \cmidrule(l){2-20}

    & 4. Remote access to Phpmyadmin   & Apache  & Y & N/A & N &\LEFTcircle & \cmark & \Circle & \Circle & \Circle & \Circle & \Circle & \Circle & \Circle & \Circle & - & - & - & - \\ \midrule

\multirow{2}{*}{2.1}  & 5. Default Mysql user/password  & Mysql & Y & 8.8 & Y & \CIRCLE             & \cmark            & \Circle            & \cmark$^*$            & \CIRCLE             & \cmark             & \CIRCLE             & \cmark              & \CIRCLE              & \cmark  & - & - & - & - \\  \cmidrule(l){2-20}

  & 6. Default Phpmyadmin passwords & Apache & Y & 8.8 & Y & \xmark   & \cmark$^*$            & \xmark            & \xmark            & \xmark             & \xmark             & \Circle             & \Circle              & \xmark              & \xmark    & - & - & - & - \\ \midrule

2.3 & 7. Sensitive information over HTTP   & Apache & Y & 8.1 & Y & \xmark             & \xmark            & \xmark            & \xmark            & \xmark             & \xmark             & \xmark              & \xmark              & \xmark              & \xmark & \xmark & \xmark & \cmark & \xmark \\ \midrule
3.2                   & 8. Store CVV in DB                                       & Webapp  & N & N/A & N/A & N/A             & N/A            & N/A            & N/A            & N/A             & N/A             & N/A              & N/A              & N/A   & N/A & N/A & N/A & N/A & N/A \\ \midrule

3.3                   & 9. Show unmasked PAN                                     & Webapp   & N   & N/A & N/A & N/A  & N/A            & N/A            & N/A            & N/A             & N/A             & N/A              & N/A              & N/A              & N/A & N/A & N/A & N/A & N/A           \\ \midrule

3.4                   & 10. Store plaintext PAN                                   & Webapp  & N  & N/A & N/A & N/A & N/A            & N/A            & N/A            & N/A             & N/A             & N/A              & N/A              & N/A              & N/A  & N/A & N/A & N/A & N/A             \\ \midrule

{3.5}  & 11. Hardcoded key for encrypting PAN   & Webapp & N & N/A & N/A & N/A  & N/A            & N/A            & N/A            & N/A             & N/A             & N/A              & N/A              & N/A & N/A  & N/A & N/A & N/A & N/A       \\ \midrule

\multirow{7}{*}{4.1} & 12. Use untrusted selfsigned cert.   & Apache    &  Y & 9.8  &  Y & \LEFTcircle             & \cmark            & \LEFTcircle            & \cmark            & \LEFTcircle             & \cmark             & \LEFTcircle              & \cmark              & \xmark              & \xmark  & - & - & - & - \\ \cmidrule(l){2-20} 

 & 13. Insecure block cipher (Sweet32)                       & Apache & Y & 7.5  &  Y & \LEFTcircle             & \cmark            & \LEFTcircle            & \cmark            & \Circle             & \Circle             & \LEFTcircle              & \cmark              & \xmark              & \xmark & - & - & - & -             \\ \cmidrule(l){2-20} 

  & 14. Expired SSL cert.                               & Apache   & Y & 5.3 &  Y & \LEFTcircle              & \cmark            & \LEFTcircle             & \cmark            & \CIRCLE             & \cmark             & \LEFTcircle              & \cmark              & \xmark              & \xmark   & - & - & - & - \\ \cmidrule(l){2-20} 

 & 15. Wrong domain names in cert.                   & Apache & Y & 5.3  &  Y & \LEFTcircle             & \cmark            & \xmark            & \xmark            & \Circle             & \Circle             & \Circle              & \Circle              & \xmark              & \xmark     & - & - & - & - \\ \cmidrule(l){2-20} 
 
  & 16. DH modulus \textless{}= 1024 Bits & Apache  & Y & 5.3 & Y & \Circle             & \cmark$^*$            & \LEFTcircle            & \cmark            & \LEFTcircle             & \cmark             & \LEFTcircle              & \cmark              & \xmark              & \xmark     & - & - & - & - \\ \cmidrule(l){2-20} 
  
 & 17. Weak Hashing in SSL cert.  & Apache   & Y & 5.3  & Y &  \LEFTcircle             & \cmark            & \xmark            & \cmark$^*$            & \xmark             & \cmark$^*$             & \LEFTcircle              & \cmark              & \xmark              & \xmark           & - & - & - & - \\ \cmidrule(l){2-20}
 
& 18. TLS 1.0 supported                                     & Apache    & Y & 3.7  & N & \CIRCLE             & \cmark            & \CIRCLE            & \cmark            & \LEFTcircle             & \cmark             & \xmark              & \xmark     &      \xmark              & \xmark        & - & - & -  & - \\ \midrule

\multirow{2}{*}{6.1}  & 19. Vulnerable OpenSSH (7.2)   & OS  & Y & 7.8  & Y &  \CIRCLE             & \cmark           & \CIRCLE            & \cmark            & \CIRCLE             & \cmark & \xmark              & \xmark             & \xmark              & \xmark      & - & - & - & - \\ \cmidrule(l){2-20}  

& 20. Vulnerable Phpmyadmin (4.8.3)                         & Apache  & Y & 6.5 & Y & \LEFTcircle             & \cmark            &  \xmark              & \xmark            &  \xmark              & \xmark             &  \xmark              & \xmark              & \xmark              & \xmark & - & - & - & - \\  
  \midrule

  \multirow{8}{*}{6.5} & 21. Sql inject in admin login   & Webapp    & Y & 9.8  & Y & \xmark             & \xmark            & \xmark            & \xmark            & \xmark             & \xmark             & \xmark              & \xmark              & \xmark              & \xmark  & \xmark & \xmark & \xmark  & \xmark  \\ \cmidrule(l){2-20} 
  
   & 22. Sql inject in customer login  & Webapp   & Y  & 9.8   & Y &  \xmark             & \xmark            & \xmark            & \xmark            & \xmark             & \xmark             & \xmark              & \xmark              & \xmark              & \xmark & \xmark & \xmark & \xmark & \cmark \\ \cmidrule(l){2-20} 
  
  & 23. Disable password retry limit                    & Webapp  & Y  & 5.3  & Y &  \xmark             & \xmark            & \xmark            & \xmark            & \xmark             & \xmark             & \xmark              & \xmark              & \xmark              & \xmark & \xmark & \xmark & \xmark & \xmark  \\ \cmidrule(l){2-20} 
  & 24. Allow passwords with len \textless 8                   & Webapp  & Y  & 5.3   & Y & \xmark             & \xmark            & \xmark            & \xmark            & \xmark             & \xmark             & \xmark              & \xmark              & \xmark  & \xmark & \xmark & \xmark & \xmark & \xmark \\ \cmidrule(l){2-20} 
  
  & 25. Javascript source integrity check  & Webapp   & Y  & 9.8 & Y & \CIRCLE             & \cmark  & \xmark            & \xmark            & \xmark             & \xmark             & \xmark              & \xmark              & \xmark              & \xmark & - & - & - & - \\ \cmidrule(l){2-20} 
  
  & 26. Don't hide program crashes                            & Webapp  & Y  & 6.5   & Y &  \xmark             & \xmark            & \xmark            & \xmark            & \xmark             & \xmark             & \xmark              & \xmark              & \xmark              & \xmark & \xmark & \xmark & \xmark & \xmark \\ \cmidrule(l){2-20} 
  
  & 27. Implant XSS                                           & Webapp  & Y & 6.1  & Y & \xmark             & \xmark            & \xmark            & \xmark            & \xmark             & \xmark             & \xmark              & \xmark              & \xmark              & \xmark & \xmark & \xmark & \xmark & \xmark  \\ \cmidrule(l){2-20}
  
  & 28. Implant CSRF                                          & Webapp    & Y  & 8.8   & Y & \LEFTcircle             & \cmark            & \xmark            & \xmark            & \xmark             & \xmark             & \xmark              & \xmark              & \xmark & \xmark & - & - & - & -  \\ \midrule
  
 \multirow{5}{*}{6.6}  & 29. Extraction of server info.   & Apache & Y & 5.3 & Y & \LEFTcircle             & \cmark            & \Circle            & \Circle            & \Circle             & \Circle             & \Circle              & \Circle              & \Circle              & \Circle      & - & - & - & -        \\ \cmidrule(l){2-20}
 
 & 30. Browsable web directories  & Apache     & Y  & 7.5 & Y & \LEFTcircle             & \cmark            &  \LEFTcircle             & \cmark            &  \LEFTcircle             & \cmark             & \LEFTcircle             & \cmark              &  \LEFTcircle             & \cmark      & - & - & - & - \\ \cmidrule(l){2-20} 
  
  & 31. HTTP TRACE/TRACK enabled                            & Apache    & Y  & 4.3     & Y & \LEFTcircle             & \cmark            &  \LEFTcircle             & \cmark            &  \CIRCLE             & \cmark             &  \LEFTcircle             & \cmark              &  \CIRCLE             & \cmark       & - & - & - &-      \\ \cmidrule(l){2-20} 
  
  & 32. phpinfo() statement is enabled   & Apache     & Y  &  5.3     & Y & \LEFTcircle             & \cmark          &  \LEFTcircle             & \cmark            & \Circle             & \Circle             & \CIRCLE              & \cmark              & \xmark              & \xmark & - & - & - & - \\ \cmidrule(l){2-20}
  
   & 33. Missing security headers in HTTP                      & Apache     & Y   & 6.1  & Y & \LEFTcircle             & \cmark            &  \LEFTcircle             & \cmark            &  \LEFTcircle             & \cmark             & \Circle              & \Circle              & \xmark              & \xmark     & - & - & - & - \\ \midrule
   
\multirow{2}{*}{8.4} & 34. Store unsalted customer passwords                     & Webapp   & N  & N/A & N/A & N/A             & N/A            & N/A            & N/A            & N/A             & N/A             & N/A              & N/A              & N/A  & N/A & N/A & N/A & N/A & N/A \\ \cmidrule(l){2-20}

& 35. Store plaintext passwords                           & Webapp     & N   & N/A     & N/A & N/A             & N/A            & N/A           & N/A            & N/A             & N/A             & N/A              & N/A              & N/A             & N/A  & N/A & N/A & N/A & N/A    \\
\midrule
               \multicolumn{6}{l?}{\textbf{Baseline: \#Vul. Detected (Total detectable: 29)}}           & \textbf{21}             & -            & \textbf{16}            & -            & \textbf{17}             & -             & \textbf{16}              & -              & \textbf{7}  & - & - & - & -  & -  \\ \midrule

\multicolumn{6}{l?}{\textbf{Certified: \#Vul. Remaining (\#Vul. detected, but no need to fix)}} &
 - & \textbf{7(0)}            & -            & \textbf{15(3)}            & -             & \textbf{18(7)}           & -      & \textbf{20(7)}               & -              & \textbf{25(4)} & - & - & -  & -  \\

\bottomrule
\end{tabular}
\end{table*}



\noindent
{\bf Iterative Test Design.} Given a PCI scanner, we carry out the evaluation in two high-level steps shown in Figure~\ref{fig:exp}. Every scanner first runs on the same baseline testbed with all the vulnerabilities built in. Then we remove a \textit{minimal} set of vulnerabilities to get the testbed certified for PCI DSS compliance. The final certified instance of the testbed may be different for different scanners, as high-quality scanners require more vulnerabilities to be fixed, having fewer remaining (undetected) vulnerabilities on the testbed.  
\begin{enumerate}
    \item {\bf Baseline Test.} We spawn a website instance where all 35 vulnerabilities are enabled (29 of them are remotely verifiable). Then we order a PCI scanning service for this testbed. During the scanning, we monitor the incoming network traffic. We obtain the security report from the scanner, once the scanning is complete.  

    \item {\bf Certified Instance Test.} After the baseline scanning, we modify the web server instance according to the obtained report. We perform all the fixes required by the PCI scanner and order another round of scanning. The purpose of this round of scanning is to identify the {\em minimal} set of vulnerabilities that need to be fixed in order to pass the PCI DSS compliance certification. 

\end{enumerate}
In summary, we perform the following steps for each scanner: {\em i)} implant vulnerabilities under each test case in the testbed, {\em ii)} run the PCI scanning, {\em iii)} fix all the vulnerabilities that the scanner mandates to fix in order to be PCI DSS compliant, {\em iv)} run the scanning again, and {\em v)} record the certified version of the testbed.

\subsection{Comparison of Scanner Performance}\label{sec:asv_eval}
We found that the security scanning capabilities vary significantly across scanners, in terms of {\em i)} the vulnerabilities they can detect and {\em ii)} the vulnerabilities they require one to fix in order to pass the certification process. Once passed, the website becomes PCI DSS compliant. The experimental results are presented in Table~\ref{table:experimental-results}. 

\noindent
\textbf{Scanner2.}
Scanner2 is the most effective PCI scanner in our evaluation, and successfully detected 21 out of the 29 externally detectable cases. The most important case that Scanner2 missed is the use of HTTP protocol to transmit sensitive information (test case 7). We fixed \textit{21} vulnerabilities in our testbed to become PCI compliant in Scanner2. Most of the fixes are intuitive, except fixing Javascript source integrity checking (Case 25) and CSRF (Case 28). We added Javascript integrity checking for scripts that are loaded from external sources (Case 25). We used a dynamic instrumentation based plugin to protect OpenCart against CSRF attacks (Case 28). This plugin instruments code for generating and checking of CSRF tokens in OpenCart forms. Sometimes, fixing one vulnerability effectively eliminates another vulnerability that Scanner2 fails to detect. For example, Scanner2 did not detect default usernames and passwords for \textit{Phpmyadmin} (Case 6); however, this vulnerability no longer exists after we disable the remote access to \textit{Phpmyadmin} (Case 4).

\noindent
\textbf{Scanner5.} In the baseline test ({\em i.e.}, when all the vulnerabilities were in place), Scanner5 detected 16 out of the 29 cases. To obtain a Scanner5 compliant version, we had to fix 13 vulnerabilities. Two of the vulnerabilities (Test cases 5 and 17) are fixed as a side effect of fixing other vulnerabilities (Test cases 2 and 12). Scanner5 failed to report the use of a certificate with the wrong hostname, which is a serious vulnerability exploitable by hackers to launch man-in-the-middle attacks. Scanner5 did not report the use of HTTP to transmit sensitive information ({\em i.e.}, login and register forms in \textit{rwycart}). Interestingly, Scanner5 detected the use of HTTP to log on to \textit{PhpMyAdmin}. In addition, Scanner5 did not report the use of scripts from external sources (Case 25).


\noindent
\textbf{Scanner1 and Scanner4.} Scanner4 uses Scanner1's scanning infrastructure for ASV scanning. So we present the experimental results of both scanners under the same column. Scanner1 detects 17 vulnerabilities. However, it only requires fixing 10 of them to be PCI DSS compliant. Some of the high and medium severity vulnerabilities are not required to fix, including remotely accessible Mysql (Case 2), certificates with wrong hostnames (Case 15), and missing security headers (Case 33).
%
%
The vulnerability of weak hashing in SSL/TLS certificates (Case 17) was fixed as a side effect of using a real certificate from Let's Encrypt (Case 12).


\noindent
\textbf{Scanner6 and Scanner3.} Scanner6 and Scanner3 are not on the approved scanning vendors (ASVs) list~\cite{pciasv} provided by the PCI council. Compared with other approved scanners, they detected a fewer number of vulnerabilities. Scanner6 detected 16 vulnerabilities, whereas Scanner3 detected 7. We fixed 9 of the vulnerabilities for Scanner6 and 3 for Scanner3 in order to be compliant. Both Scanner6 and Scanner3 detected remotely accessible Mysql (Case 2), but do not require us to fix them. Scanner3 missed all the SSL/TLS and certificate related vulnerabilities (Test cases 12-18), while Scanner6 detected most of them. However, Scanner6 did not require us to fix certificates with wrong hostnames (Test case 15). We cannot conclude that unapproved scanners perform worse than approved scanners, due to the small sample size.

\noindent
\textbf{A Case Study of False Positives.} During our experiment, we find Scanner2 produced a false positive under the SQL injection test. Scanner2 recently incorporated an experimental module to find blind SQL detection, by sending specially crafted parameters to the web server. If the server returns different responses, then it determines that the server has accepted and processed the parameter ({\em a.k.a} vulnerable). However, this detection procedure fails on a common e-commerce scenario: supporting multiple currencies. OpenCart allows users to select the currency for a product. If a currency is clicked, it updates the currency of the current page. The server records all the parameters of the current page under a hidden field so that it can recreate the page later (Listing~\ref{paraminjection}). Note that Scanner2's specially-crafted parameters are also recorded, which makes Scanner2 believe that there exists a difference in the output under different values of the parameter, which is actually a false positive. Nevertheless, we fixed it by changing the {\sc BuggyCart} code to be certified by Scanner2.

\vspace{-0.1in}
\definecolor{backcolour}{rgb}{0.95,0.95,0.92}
\lstset{
    backgroundcolor=\color{backcolour},  
    basicstyle=\footnotesize,
    breakatwhitespace=false,         
    language=html,
  caption={The difference in the output after injecting a parameter named \textit{name} with an empty value ``'' vs. ``yy''.},
  label=paraminjection
}
\begin{lstlisting}
<input type="hidden" name="redirect"
value="http://www.rwycart.com/upload
/index.php?...&amp;product_id=49&amp;name="/>

-------- vs --------

<input type="hidden" name="redirect"
value="http://www.rwycart.com/upload
/index.php?...&amp;product_id=49&amp;name=yy"/>
\end{lstlisting}

\noindent
\textbf{Network Traffic Analysis.} We collected the incoming network requests from each of the scanners using the access log of our testbed. During the baseline experiment, Scanner5, Scanner6 and Scanner3 sent 23,912, 39,836, and 31,583 requests, respectively and finished within an hour. Scanner4 and Scanner1 sent 147,038 requests and took more than 3 hours to finish. Scanner2 sent 64,033 requests within 2.5 hours. The reason why we received such a high traffic volume is that the PCI scanners were attempting to detect vulnerabilities such as XSS, SQL injection that require intensive URL fuzzing, crawling and parameter manipulations. This confirms that the PCI scanners have at least attempted to detect such vulnerabilities but were just unsuccessful.  



\subsection{Impacts of Premature Certification}\label{sec:securityissues}  

Some scanners choose to simply report vulnerabilities without marking the e-commerce website as non-compliant. Below, we discuss the security consequences of premature certifications. Some of the incomplete scanning and premature certification issues can be prevented, if the scanners follow the ASV guidelines~\cite{DBLP:books/asv/guideline}.  

\noindent
{\bf Network Security Threats.} According to the ASV scanning guideline, SSL/TLS vulnerabilities (Test cases 12--17) should lead to automatic failure of certification, which is clearly necessary due to the man-in-the-middle threats. Only Scanner2 detected all these cases. Scanner3 does not detect any of these SSL/TLS vulnerabilities.
In addition, a website should be marked as non-compliant if sensitive information is communicated over HTTP (Test case 7). However, none of the ASV scanners detected this issue in our testbed. This vulnerability can be avoided by configuring the server to automatically redirect all the HTTP traffic to HTTPS. 
Because none of the 6 scanners detected this vulnerability, it is likely that this HTTP issue exists on real-world e-commerce websites. Our later evaluation on 1,203 websites that process online payment shows 169 of them do not redirect their HTTP traffic to HTTPS (Section~\ref{sec:ourchecker}).


Our Test case 2 embeds a database access vulnerability, allowing the database to be accessible from the Internet. All the scanners detected this vulnerability. However, only Scanner2 and Scanner5 mark this issue as an automatic failure ({\em i.e.}, non-compliant). The other scanners report it as ``low/information'', not as a required fix, despite the ASV scanning guideline~\cite{DBLP:books/asv/guideline} recommends that to be marked as non-compliant. Our evaluation later on websites that accept payment card transactions shows that 59 out of 1,203 websites opened the Mysql port (3306) to the Internet (Section~\ref{sec:ourchecker}).

\noindent
{\bf System Security Threats.} The ASV scanning guideline~\cite{DBLP:books/asv/guideline} suggests to test and report vulnerable remote access software. 4 out of the 6 scanners detected vulnerable OpenSSH software  (Test case 19).  Under Test case 20, only Scanner2 detected vulnerable {\em phpmyadmin}, while others failed. Although all scanners noticed the Test case 29 (extracted server information), only Scanner2 required a fix for compliance.
The ASV scanning guideline~\cite{DBLP:books/asv/guideline} also recommends reporting automatic failure if browsable web directories are found (Test case 30). All scanners detected this vulnerability. Scanner6 detected missing security headers (Test case 33), but did not ask us to fix it, while Scanner3 failed to detect it.

\noindent
{\bf Web Application Threats.} The scanners' performance is particularly weak for this category.
Out of the 8 test cases, only 2 were detected by Scanner2 (tampered Javascript and CSRF). None of the cases was detected by other PCI scanners. 






\subsection{Evaluation of Website Scanners}
\label{sec:eval-website-scanners}
The above results suggest that some web application vulnerabilities are difficult to detect. The follow-up question is, {\em can specialized website scanners detect these vulnerabilities?} To answer this question, we ran four website scanners on our {\sc BuggyCart} testbed, including two commercial ones (from Scanner2 and Scanner5) and two open source scanners (w3af~\cite{w3aflink} and ZAP~\cite{zaplink}). w3af and ZAP are state-of-the-art open source web scanners, are actively being maintained and are often used in academic research~\cite{DBLP:conf/uss/DoupeCKV12, DBLP:conf/dimva/DoupeCV10, DBLP:journals/wicomm/RamosVL18}. The two commercial web scanners are from reputable companies. Scanner2W and Scanner2 are from the same company, where the website scanner is marketed as a different product from PCI scanner. It is the same for Scanner5W and Scanner5. 

%

We conducted the baseline test for the four website scanners. Note that these web scanners do not produce certificates. The results are shown in the last four columns of Table~\ref{table:experimental-results}. Since they are website scanners, we only expect them to cover web application vulnerabilities (Test case 7, 21, 22, 23, 24, 26, 27). Unfortunately, none of the commercial scanners detect these web application vulnerabilities. W3af reported the use of HTTP protocol to communicate sensitive information (case 7), but missed other web application vulnerabilities. ZAP detected the SQL injection vulnerability in the customer login page (case 22), but missed the SQL injection vulnerability in the admin login page (case 21). Noticeably, ZAP also missed the XSS vulnerability we implanted (case 27).




\noindent
{\bf Summary of Testbed Findings.} The detection capabilities vary significantly across scanners. Our experiments show that 5 out of 6 PCI scanners are not compliant with the ASV scanning guidelines~\cite{DBLP:books/asv/guideline} by ignoring detected vulnerabilities and not making them ``automatic failures''. 
Most of the common web application vulnerabilities ({\em e.g.}, SQL injection, XSS, CSRF) are not detected by the 6 PCI scanners (only Scanner2 detected CSRF), despite the requirements of the PCI guidelines. Out of the 4 website scanners, only ZAP detected one of the two SQL injection cases. 

Admittedly, black-box detection of vulnerabilities such as XSS and SQL injection is difficult. Typical reasons for missed detection are {\em i)} failure to locate the page due to incomplete discovery and/or {\em ii)} that detection heuristics are limited or easily bypassed. In our testbed, SQL injection vulnerabilities (21, 22) are placed in the login pages. CSRF vulnerabilities are present in all forms. The scanners we tested used web crawling with URL fuzzing to detect hidden pages, URLs, and functions. Often, we are unable to pinpoint the exact reasons why the tools fail in these cases. 
Novel detection techniques, such as guided fuzzing~\cite{DBLP:conf/uss/DoupeCKV12} and taint tracking~\cite{DBLP:conf/ndss/SteffensRJS19}, have been proposed by the research community. Future work is needed to evaluate their applicability in the specific PCI context.


%
%
%

\section{Measurement of Compliant Websites}
\label{sec:ourchecker}

The alarming security deficiencies in how PCI scanners conduct the compliance certification motivate us to ask the following questions: {\em How secure are e-commerce websites?} {\em What are the main measurable vulnerabilities in e-commerce websites?} As such we designed another set of real-world experiments where we aim to measure the security of e-commerce websites with respect to the PCI DSS guideline. To do so, we need to address several technical questions.  

\noindent
{\bf What Tools to Use?} The key enabler of this measurement is a new tool we developed named \checker{}. We use basic Linux tools ({\em e.g.}, \textit{nc}, \textit{openssl}) and Java net URL APIs to implement the system. Below, we focus on the key design concepts of \checker{} in order to work with {\em live websites}. 

\noindent
{\bf What Security Properties to Check?} A key requirement of this experiment to make sure that we do not disrupt or negatively impact websites being tested. Out of the 29 externally verifiable cases in Table~\ref{table:experimental-results}, we choose a subset of 17 cases for this experiment, as shown in Table~\ref{table:checkerresult}. The sole reason of selecting these cases for \checker{} is that we could implement these tests in a non-intrusive manner, leaving a minimum footprint, {\em i.e.}, having a minimum impact on the servers. We categorize these cases to high, medium and low severity based on {\em i)} the attacker's gain and {\em ii)} the attack difficulty. Cases that are immediately exploitable by any arbitrary attacker to cause large damages are highly severe, for example, the use of default passwords (Test case 5), insecure communications (Test cases 7, 12, 13, 16, 17), vulnerable remote access software (Test case 19), browsable web directories (Test case 22), and supporting HTTP TRACE method (Test case 23). Cases that substantially benefit any arbitrary attacker but require some efforts to exploit are marked as medium severity, {\em e.g.}, test cases 2, 14, 15, 25, 29, and 33. For example, scripts loaded from external sources can steal payment card data (Test case 25), but attackers need to craft the malicious scripts~\cite{krebsonsecurity:whois}. Low-risk issues are marked as low severity (Test case 3, 18). The categories are consistent with Table~\ref{table:experimental-results} as high and medium severity cases correspond to ``must-fix'' vulnerabilities. The two low-severity cases are not required to be fixed to be PCI-compliant.  

\noindent
{\bf Implementing \checker{}.} Our goal is to minimize the number of requests that \checker{} issues per test case, while maximizing the test case coverage. It involves a collection of lightweight heuristics that merge multiple tests into a single request. For example, for most of the HTTP-related tests, we reuse a single response from the server. Test cases 25, 29, and 33 are covered and resolved by one single HTTP request to retrieve the main page and analyzing the response header. Test cases 12--18 are covered by one certificate fetching. For case 30 (browsable directories enabled) \checker{} conducts a code-guided probe and avoids crawling web pages. It discovers file paths in the code of the landing page and then probes the server with requests for accessing path prefixes. The implementation details are given in the Appendix.

\begin{table}[t]
\caption{Number of e-commerce websites that have at least one vulnerability and those that have at least one ``must-fix'' vulnerability. In total, 1,203 sites are tested including 810 sites chosen from different web categories, and 393 sites chosen from different Alexa ranking ranges.}\label{table:one-or-2-vuls}
\small
\vspace{-0.1in}
\begin{tabular}{@{}llcc@{}}
\toprule
 \multicolumn{2}{c}{E-commerce Websites}  & \multicolumn{2}{c}{\# of Vulnerable Websites} \\ \cmidrule(l){3-4}
  &   & Must-fix Vul. & All Vul. \\ \midrule
\multirow{10}{*}{Category (810)} & Business (122)   & 106 & 113 \\
  & Shopping (163)   & 135 & 143         \\
  & Arts (78)     & 66 & 76         \\
  & Adults (65)    &  61 & 65        \\
  & Recreation (84) & 70 & 75         \\
  & Computer (57) & 53 & 56         \\
  & Games (42) & 38 & 42         \\
  & Health (60) & 54 & 55         \\
  & Home (102) & 82 & 93        \\
  & Kids \& Teens (37) & 31 & 36         \\
    \midrule 
\multirow{2}{*}{Ranking (393)}  & Top (288)  & 235 & 277  \\
  & Bottom (105)    &  100 & 104         \\
      \midrule
 Total (1,203)  &  {}   & 1,031 (86\%) & 1,135 (94\%)        \\ 
\bottomrule
\end{tabular}
\vspace{-3mm}
\end{table}

\newcolumntype{?}{!{\vrule width 1.5pt}}
\begin{table*}[!ht]
\caption{Testing results on 1,203 real-world websites that accept payment card transactions as of May 3, 2019. We reuse the index numbers of the test cases from Table~\ref{table:experimental-results}.  }\label{table:checkerresult}
\vspace{-0.1in}
\renewcommand{\aboverulesep}{0pt}
\renewcommand{\belowrulesep}{0pt}
\renewcommand{\arraystretch}{1.2}
\setlength{\tabcolsep}{0.5em}

\footnotesize

\begin{tabular}{@{}lll|llllllllll|ll?l@{}}
\toprule
\multirow{2}{*}{Reqs.} & \multirow{2}{*}{Test Cases} & \multirow{2}{*}{Severity} & \multicolumn{10}{c}{Category (810)} & \multicolumn{2}{|l?}{Ranking (393)} & Total (1,203)\\ 
\cmidrule{4-15} 
&  & & {\begin{tabular}[c]{@{}l@{}} Biz.\\ (122)\end{tabular}}  & {\begin{tabular}[c]{@{}l@{}} Shop.\\ (163)\end{tabular}} & {\begin{tabular}[c]{@{}l@{}} Arts \\ (78)\end{tabular}} & {\begin{tabular}[c]{@{}l@{}} Adlt.\\ (65)\end{tabular}} & {\begin{tabular}[c]{@{}l@{}} Recr.\\ (84)\end{tabular}} &
{\begin{tabular}[c]{@{}l@{}} Comp.\\ (57)\end{tabular}} &
{\begin{tabular}[c]{@{}l@{}} Game.\\ (42)\end{tabular}} &
{\begin{tabular}[c]{@{}l@{}} Hlth.\\ (60)\end{tabular}} &
{\begin{tabular}[c]{@{}l@{}} Home.\\ (102)\end{tabular}} &
{\begin{tabular}[c]{@{}l@{}} Kids.\\ (37)\end{tabular}} &
\multicolumn{1}{c}{\begin{tabular}[c]{@{}l@{}} Top \\ (288)\end{tabular}} & \multicolumn{1}{c?}{\begin{tabular}[c]{@{}l@{}} Btm. \\ (105)\end{tabular}} & \\ \cmidrule{1-16}
\multirow{2}{*}{1.2}   & 2. Mysql port (3306) detection   & Medium  & 3 & 6 & 4 & 2 & 6 & 2 & 3 & 2 & 4 & 0 & 0 & 27 & 59 (5\%) \\
& 3. OpenSSH available & Low & 6 & 15 & 11 & 4 & 13 & 6 & 7 & 8 & 12 & 1 & 6 & 27 & 116 (10\%) \\ \cmidrule{1-16}

2.1 & 5. Default Mysql user/passwd & High & 0 & 0 & 0 & 0 & 0 & 0 & 0 & 0 & 0 & 0 & 0 & 0 & 0 (0\%) \\ \cmidrule{1-16}

2.3 & 7. Sensitive info over HTTP & High & 12 & 10 & 12 & 10 & 17 & 10 & 8 & 6 & 10 & 5 & 47 & 22 & 169 (14\%)                        \\ \cmidrule{1-16}

\multirow{7}{*}{4.1} & 12. Selfsigned cert presented   & High & 0 & 0 & 3 & 0 & 1 & 0 & 0 & 1 & 1 & 0 & 0 & 3 & 9 (1\%) \\
 & 13. Weak Cipher Supported   & High  & 0 & 0 & 0 & 0 & 0 & 0 & 0 & 0 & 0 & 0 & 0 & 0 & 0 (0\%) \\
 & 14.  Expired cert presented  & Medium & 0 & 0 & 2 & 0 & 2 & 0 & 0 & 1 & 0 & 0 & 0 & 2 & 7 (1\%) \\

  & 15. Wrong hostname in cert & Medium & 3 & 1 & 3 & 0 & 6 & 2 & 0 & 2 & 4 & 1 & 0 & 10 & 32 (3\%) \\
 & 16. Insecure Modulus    & High  & 0 & 0 & 0 & 0 & 0 & 0 & 0 & 0 & 0 & 0 & 0 & 1 & 1 (0.1\%) \\
 & 17. Weak hash in cert   & High & 0 & 0 & 0 & 0 & 0 & 0 & 0 & 0 & 0 & 0 & 0 & 0 & 0 (0\%)\\
  & 18. TLSv1.0 Supported & Low   & 67 & 73 & 53 & 42 & 41 & 40 & 28 & 30 & 67 & 16 & 216 & 71 & 744 (62\%) \\ \cmidrule{1-16}

\multirow{1}{*}{6.1} & 19. OpenSSH vulnerable & High   & 6 & 14 & 11 & 4 & 13 & 6 & 6 & 8 & 11 & 1 & 6 & 26 & 112 (9\%) \\ \cmidrule{1-16}

6.5  & 25. Missing script integrity check & Medium & 92 & 109 & 54 & 44 & 44 & 32 & 27 & 42 & 66 & 21 & 154 & 75 & 760 (63\%) \\ \cmidrule{1-16}

 \multirow{4}{*}{6.6}& 29. Server Info available   & Medium & 26 & 34 & 17 & 17 & 22 & 15 & 17 & 17 & 25 & 11 & 33 & 22 & 256 (21\%) \\
 
  & 30. Browsable Dir Enabled  & High &0 & 0 & 0 & 0 & 0 & 0 & 0 & 0 & 0 & 0 & 0 & 0 & 0 (0\%) \\
 
 & 31. HTTP TRACE supported   & High & 6 & 4 & 3 & 3 & 2 & 5 & 2 & 2 & 6 & 0 & 4 & 6 & 43 (4\%) \\
  & 33. Security Headers missing  &  Medium  & 18 & 38 & 9 & 12 & 14 & 21 & 9 & 7 & 14 & 7 & 114 & 13 & 276 (23\%) \\ \bottomrule
\end{tabular}


\end{table*}

\noindent
{\bf How to Determine Whether a Website is PCI Compliant?} It is not easy to directly confirm whether a website is DSS compliant or not, unless the website actively advertises this information. While some cloud and service providers ({\em e.g.}, Google Cloud~\cite{google-cloud-pci}, Amazon Connect~\cite{amazon-connect-pci}, Shopify~\cite{shopify-pci}, and Akamai~\cite{akamai-pci}) advertise their PCI compliance status, not all of them disclose such information. However, as e-commerce websites need to show their DSS compliance in order to work with acquirer banks (described in Section~\ref{sec:back}), it is reasonable to assume that most websites we evaluated have successfully passed the external scanning.





\noindent
{\bf  Website Selection.} We use two different ways to select websites to increase diversity. First, we downloaded 2,000 Alexa top websites under 10  categories (200 websites per category) to observe security compliance differences based on categories. In Table~\ref{table:one-or-2-vuls}, we show the category-wise breakdown. 
Among them, we manually identified 810 websites that make payment card transactions. This step is time-consuming and usually requires manually visiting multiple pages ({\em e.g.}, one needs to visit multiple pages to get to the payment page on \textit{nytimes.com}). 
Second, to cover websites of different popularity levels, we further select the top 500 and bottom 500 websites (1,000 in total) from Alexa top 1 million website list.  We found 288 websites from the top list and 105 websites from the bottom list that accept payment card information (and do not overlap with the previous 811 websites).
In total, 1,203 payment-cards-taking websites are selected for scanning by \checker{}. 




\noindent
{\bf  Findings of E-commerce Website Compliance.} 
68 websites fully passed our \checker{} test, including the aforementioned cloud providers (Google Cloud, Amazon Connect, Shopify). Our results also confirm that a number of actively operating websites do not comply with the PCI Data Security Standard.  As shown in Table~\ref{table:one-or-2-vuls}, out of the 1,203 websites, 1,135 (94\%) have at least one vulnerability. More importantly, 1,031 (86\%) sites have at least one vulnerability that belong to the ``must-fix'' vulnerabilities which should have disqualified them as non-compliant. Among them, 520 (43\%) sites even have two or more must-fix vulnerabilities. 

Then as shown in Table~\ref{table:checkerresult}, the shopping category has the lowest percentage (87\%) of vulnerable websites, while all other categories have a percentage of over 90\%. We found several types of high-risk and medium-risk vulnerabilities, including leaving the Mysql port (3306) open, using self-signed or expired certificates, wrong hostnames in the certificate, enabling HTTP TRACE method, and using vulnerable OpenSSH (7.5 or earlier).
 Supporting TLS v1.0 (low-risk level) is another most common vulnerability we detected (Test case 18), likely due to the need for backward compatibility. SSLv3.0 and TLSv1.0 are known to have multiple man-in-the-middle vulnerabilities~\cite{DBLP:conf/ccs/AdrianBDGGHHSTV15} and the PCI standard recommends that all web servers and clients must transition to TLSv1.1 or above. 

The vulnerabilities in these websites suggest the PCI scanners used by the websites are inadequate and failed to detect the vulnerabilities during the certification scans. Another possibility is that the acquiring banks did not sufficiently examine the merchants' quarterly security reports, allowing merchants to operate without sending adequate security reports to banks as required. 

\noindent
{\bf Vulnerable Websites.} Below, we highlight some interesting findings without explicitly mentioning the names of vulnerable sites.

\noindent
{\em Mysql open ports.} 59 websites expose the MySql service for remote access. For example, two Slovenian websites that sell healthcare products and car components and a Russian website that sells furnaces and stoves all have this vulnerability. We did not detect any use of default user (root) or no password.


\noindent
{\em Insecure certificates (self-signed, expired, and insecure modulus).} The use of certificates with wrong hostnames (Figure~\ref{fig:wronghostname} in Appendix) is an issue that appears in 3\% of the websites. 
For some websites, the root cause is not properly configuring HTTPS. For example, one website accepts payment for donations. Since it does not correctly set up HTTPS, it uses a default certificate\footnote{A self-signed certificate comes with the webserver installation.} for HTTPS (Figure~\ref{fig:selfsigned} in Appendix). In some cases, the websites use HTTPS for payment only while other sensitive content ({\em i.e.}, items and the cart) are still sent over HTTP. Because the original domain is not properly configured to use HTTPS, it presents the default expired certificate (Figure~\ref{fig:expired} in Appendix). 



\begin{table}[t]
\caption{Comparison between \checker{} and the {\tt customized w3af} on 100 randomly chosen websites and the Buggycart testbed. We report the number of vulnerable websites detected and the false positives (FP) among them. }\label{result:pci-checker-vs-w3af}
\vspace{-0.05in}
\small 
\begin{tabular}{@{}lll|ll@{}}
\toprule
\multirow{2}{*}{Vulnerabilities}   & \multicolumn{2}{c|}{100 Random websites} & \multicolumn{2}{c}{Buggycart} \\ \cmidrule(l){2-5} 
                                   & Ours (FP)          & w3af (FP)         & Ours       & w3af       
\\ \cmidrule{1-5}
2. Mysql port (3306) detection     & 5 (0)                   & 0 (0)            &  \cmark   &  \xmark \\
3. OpenSSH available               & 10 (0)                   & 0 (0)            &  \cmark   &  \xmark \\
5. Default Mysql user/pass         & 0 (0)                    & 0 (0)            &  \cmark   &  \xmark \\
7. Sensitive info over HTTP        & 12 (0)                  & 27 (17)            &  \cmark   &  \cmark \\
12. Selfsigned cert presented      & 2 (0)                   & 2 (0)             &  \cmark   &  \cmark \\
13. Weak Cipher Supported          & 0 (0)                  & 0 (0)           &  \cmark   &  \xmark \\
14. Expired cert presented         & 0 (0)                   & 3 (3)             &  \cmark   &  \cmark \\
15. Wrong hostname in cert         & 3 (0)              & 2 (1)             &  \cmark   &  \cmark \\
16. Insecure Modulus               & 0 (0)                   & 0 (0)              &  \cmark   &  \xmark \\
17. Weak hash in cert              & 0 (0)                   & 0 (0)         &  \cmark   &  \xmark \\
18. TLSv1.0 Supported              & 63 (0)                  & 0 (0)            &  \cmark   &  \xmark \\
19. OpenSSH vulnerable             & 10 (0)                  & 0 (0)            &  \cmark   &  \xmark \\
25. Missing script integrity check & 72 (1)                  & 55 (10)            &  \cmark   &  \cmark \\
29. Server Info available          & 19 (0)                  & 81 (62)         &  \cmark   &  \cmark \\
30. Browsable Dir Enabled          & 0 (0)                    & 0 (0)               &  \cmark   &  \xmark \\
31. HTTP TRACE supported           & 6 (0)                  & 6 (6)        &  \cmark   &  \cmark \\
33. Security Headers missing       & 30 (0)                  & 0 (0)     &  \cmark   &  \xmark  \\ \bottomrule
\end{tabular}
\vspace{-0.12in}
\end{table}
 

\noindent
{\bf Comparison with Existing Tool.} Finally, we experimentally compared \checker{} with the state-of-the-art web scanner. Note that existing scanners typically have aggressive pentesting components that are not suitable to test live websites. For this experiment, we choose w3af and have to adapt it to a ``non-intrusive low-interactive" version. More specifically, we modify w3af to 1) block intrusive 
tests ({\em e.g.}, XSS, SQL injections), 2) disable URL fuzzing, and 3) disable the liveliness testing. For scalability, we also utilized w3af's programmable APIs (w3af\_console) to discard the graphic user interface. We call this version as {\em customized w3af}. For comparison, we ran \checker{} and the customized w3af on 100 websites random from the 1203 sites (in Table~\ref{table:checkerresult}). For reference, we also ran both tools on our {\sc BuggyCart}. 

The results are shown in Table~\ref{result:pci-checker-vs-w3af}. First, we observe that our system outperforms w3af on Buggycart by detecting all the vulnerabilities. Second, on the 100 real-world websites, our system also detected more {\em truly vulnerable} websites. Even though w3af flagged more websites ({\em e.g.}, Test case 7, 29), manually analysis shows that a large portion of the alerts are false positives. For example, under Test case 7, w3af flags a website if Port 80 is open, while \checker{} reports a website only if the request is not automatically redirected to Port 443 (HTTPS). This design of w3af produces 17 false positives. Under Test case 15, w3af flags a website that uses the certificate for its subdomains (which is not a violation). For Test case 29, w3af flags websites that expose non-critical information whereas we only flag the exposure of exploitable information ({\em e.g.}, server and framework version numbers). Note that among all vulnerabilities, we only have one FP under Test case 25. This website is flagged by \checker{} for loading Javascript from a third-party domain without an integrity check. Manually analysis shows that the third-party domain and the original website are actually owned by the same organization. Technically, such information is beyond what \checker{} can collect.

\section{Disclosure and Discussion}
\label{sec:discuss}

\smallskip
\noindent
{\bf Responsible Disclosure.} We have fully disclosed our findings to the PCI Security Standard Council. In May 2019, we shared our paper with the PCI SSC, and successfully got in touch with an experienced member of the Security Council. Through productive exchanges with them, we gained useful insights. 
{\em i)} The Security Council shared a copy of our paper to the dedicated companies that host the PCI certification testbeds, who are now aware of our findings; 
{\em ii)} Preventing scanners from {\em gaming the test} is one of their priorities, for example, by constantly updating their testbeds and changing the tests; 
{\em iii)} Low interaction constraints make it difficult to test some vulnerabilities externally (which we also experienced and aimed to address in our work); 
{\em iv)} The Security Council routinely removes scanners from the ASV list or warns scanners based on the feedback sent by ASV consumers;
{\em v)} Their testbeds exclude vulnerabilities whose CVSS scores are lower than 4; 
{\em vi)} Payment brands and acquirer banks need efficient (and automatic) solutions to inspect PCI DSS compliance reports. 
Insights {\em ii)}, {\em iii)}, and {\em vi)} present interesting research opportunities.
In addition, we are in the process of contacting vulnerable websites. Some notifications have been sent out to those that failed test case 2 (open Mysql port) or 19 (vulnerable OpenSSH). Incidentally, we found a few websites have already fixed their problems, for example Netflix upgraded the vulnerable SSH-2.0-OpenSSH\_7.2p2 (current Netflix.com server does not show a version number).

\noindent
{\bf Is Improving PCI Certification a Practical Task?}
From the economics point of view, the concept of for-profit security certification companies may seem like an oxymoron. Intuitively, a scanning vendor might make more money if its scanner is less strict, allowing websites to easily pass the DSS certification test. On the contrary, a company offering rigorous certification scanning might lose customers when they become frustrated from failing the certification test. Phenomena with misaligned incentives widely exist in many security domains ({\em e.g.}, ATM security, network security)~\cite{Anderson-Science-2006}. 
%
Fortunately, unlike the decentralized Internet, PCI security is centrally supervised by the PCI Security Council. Thus, the Council, governing the process of screening and approving scanner vendors, is a strong point of quality control. The enforcement can be strengthened through technical means. Thus, improving the PCI security certification, unlike deploying Internet security protocols~\cite{routing-misaligned-incentives-2018}, is a practical goal that is very reachable in the near future. 

\noindent
{\em Gaming-resistant Self-evolving Testbeds and Open-Source PCI Scanners.} 
A testbed needs to constantly evolve, incorporating new types of vulnerabilities and relocating existing vulnerabilities over time. A fixed testbed is undesirable, as scanners may gradually learn about the test cases and trivially pass the test without conducting a thorough analysis. 
Automating this process and creating self-evolving testbeds are interesting open research problems.

Competitive open-source PCI/web scanners from non-profit organizations could drive up the quality of commercial vendors, forcing the entire scanner industry to catch up, and providing alternative solutions for merchants to run sanity check on their services. Currently, there are not many high-quality, open-source and deployable web scanners; {\tt w3af} and {\tt ZAP} are among the very few available. 


\noindent
{\em Automate the Workload at Payment Brands and Acquirer Banks.} 
Payment brands and acquirer banks are the ultimate gatekeepers in the PCI DSS enforcement chain. Manually screening millions of scanning reports and questionnaires every quarter is not efficient (and is likely not being done well in practice). Indeed, our real-world experiments suggest that the gatekeeping at the acquirer banks and payment brands appears weak. Thus, automating the report processing for scalable enforcement is urgently needed.

\noindent
{\bf Scanning vs. Self-assessment Questionnaires.}
There are four major types of Self-assessment Questionnaires or SAQs (A to D)~\cite{DBLP:books/saq/guideline}. The different SAQs are designed for different types of merchants, as illustrated in Figure~\ref{fig:saq_chart1} in the Appendix. In SAQs, all the questions are close ended, {\em i.e.}, multiple choices. For a vast majority of the merchants, the current compliance checking largely relies on the trust of a merchant's honesty and capability of maintaining a secure system. This observation is derived from our analysis of the 340 questions in the self-assessment questionnaire (SAQ) D-Mer, which is an SAQ designed for merchants that process or store cardholder data. Consequently, it is the most comprehensive questionnaire.


We manually went through all the questions the in Self-Assessment Questionnaire (SAQ) D-Mer and categorized them into the five major groups, {\em network security}, {\em system security}, {\em application security}, {\em application capability}, and {\em company policies}. 
271 of the 340 questions fall under the category of company policies and application capability, where none of them can be automatically verifiable by an external entity ({\em e.g.}, ASV/web scanners). Only 31 out of the 69 questions on network, system and application security are automatically verifiable by a PCI scanner. 

\noindent
{\bf Legal Consequences of Cheating in PCI Certification.} The PCI DSS standard is not required by the U.S. federal law. Some state laws do refer to PCI DSS ({\em e.g.}, Nevada, Minnesota, Washington)~\cite{washington}, stating that merchants must be PCI compliant. However, there is no mentioning about any legal consequences of cheating in the PCI DSS certification process. Thus, it appears that being untruthful when filling out the self-assessment questionnaire would not have any direct legal consequences. The only potential penalty would be an ``after effect''. For example, a merchant may be fined by the card brand if a data breach happens due to its non-compliance~\cite{pcilegal}.

\noindent
{\bf Limitations.} Our work has a few limitations. First, we only tested 6 PCI scanners and 4 web scanners. Given the high expense to order PCI and web scanning, it is unlikely that such an experiment can truly scale up. For PCI scanning, we have tried to increase the diversity of scanner selection by selecting from different price ranges. The website scanners are added to further increase diversity. Second, our paper primarily focuses on the PCI compliance certification of e-commerce websites. Although we did not evaluate the compliance of {\em banks} (which report to card brands), we argue that it is the same set of the approved PCI scanners that provide the compliance reports for both merchants and banks. The problem revealed in our study should be generally applicable. Third, we did not test vulnerabilities that are not yet covered by the current Data Security Standards (DSS). Future work can further study the comprehensiveness of DSS. Finally, in Section~\ref{sec:ourchecker}, we only tested 1,203 e-commerce websites because it requires manual efforts to {verify whether a website accepts payment card information}. It is difficult to automate the verification process since one often needs to register an account and visit many pages before finding the payment page. We argue that our experiment already covers websites from various categories and ranking ranges, which is sufficient to demonstrate the prevalence of the problem.

\section{Related Work}


\noindent
\textbf{Website Scanning.} 
The detection of web application vulnerabilities has been well studied by researchers~\cite{DBLP:conf/ndss/SteffensRJS19, DBLP:conf/sp/BauBGM10, DBLP:conf/uss/DoupeCKV12}. In~\cite{DBLP:conf/sp/BauBGM10,DBLP:conf/dsn/VieiraAM09}, authors measured the performance of several black-box web scanners and reported a low detection rate for XSS and SQL injection attacks. The main challenge is to exhaustively discover various web-app states by observing the input/output patterns. Duchene \textit{et al.}~\cite{DBLP:conf/codaspy/DucheneRRG14} proposed an input fuzzer to detect XSS vulnerabilities. Doup{\'{e}} \textit{et al.}~\cite{DBLP:conf/uss/DoupeCKV12} proposed to guide fuzzing based on the website's internal states. In~\cite{DBLP:conf/ndss/PellegrinoB14}, authors proposed a black-box method to detect logical flaws using network traffic. In~\cite{DBLP:conf/ndss/SteffensRJS19}, authors used a taint-tracking based detection of XSS vulnerabilities at the client-side. In~\cite{DBLP:conf/ccs/PellegrinoJ0BR17}, authors used dynamic execution trace-based behavioral models to detect CSRF vulnerabilities. Although most defenses against XSS and SQL inject attacks prescribe input sanitization~\cite{DBLP:conf/sp/BalzarottiCFJKKV08, DBLP:conf/uss/HooimeijerLMSV11, DBLP:conf/popl/LivshitsC13}, in~\cite{DBLP:conf/ccs/DoupeCJPKV13}, authors proposed an application-agnostic rewrite technique to differentiate scripts from other HTML inputs. We argue that similar research efforts could make a positive impact to the PCI community by (1) producing and releasing high-quality open-sourced tools; and (2) customizing a non-intrusive version of the tool for testing production websites in the PCI DSS context.

\noindent
\textbf{Proactive Threat Measurements.} Honeypots~\cite{DBLP:conf/uss/Provos04, DBLP:conf/leet/Nazario09} are useful to collect empirical data on attackers (or defenders).  In~\cite{DBLP:conf/ccs/HanKB16}, authors measure attack behaviors by deploying vulnerable web servers waiting to compromised. In~\cite{oest2019phishfarm}, authors deployed phishing websites to measure the timeliness of browsers' blacklist mechanisms. In~\cite{DBLP:conf/www/CanaliBF13}, authors measure the capability of the web hosting providers to detect compromised websites by deploying vulnerable websites within those web hosting services. Our testbed can be regarded as a specialized honeypot to assess the capability of PCI scanners. 

\noindent
\textbf{Physical Card Frauds.} Payment card frauds at ATM or point-of-sale (POS) machines have been studied for decades~\cite{DBLP:conf/ccs/Anderson93, DBLP:conf/uss/DrimerM07, DBLP:conf/sp/MurdochDAB10, DBLP:conf/sp/BondCMSA14, DBLP:journals/cacm/AndersonM14, DBLP:conf/uss/ScaifePT18, DBLP:conf/sp/ScaifePVZTA18}. Most of these frauds occur due to stealing payment card information during physical transactions~\cite{DBLP:conf/ccs/Anderson93, krebsonsec:insetSkimmerCam}, and cloning magnetic stripe cards~\cite{DBLP:conf/uss/ScaifePT18,DBLP:conf/sp/ScaifePVZTA18}. EMV cards are known to be resistant to card cloning, but are vulnerable to tempered terminals~\cite{DBLP:conf/uss/DrimerM07}, or due to protocol-level vulnerabilities~\cite{DBLP:conf/sp/MurdochDAB10} and implementation flaws~\cite{DBLP:conf/sp/BondCMSA14}. Recently, researchers proposed mechanisms to detect magnetic card skimmers~\cite{DBLP:conf/uss/ScaifePT18, gaspump}.



\noindent
\textbf{Digital Card Frauds.} In the {\em online} setting, the danger of using \textit{magnetic-stripe-like transactions} is known for years~\cite{geminiy:creditcardfraud, krebsonsec:howtogetCVV}. Various methods ({\em e.g.}, 3D-Secure~\cite{DBLP:books/emv/3ds}, Tokenization framework~\cite{DBLP:books/emv/tokenisation}) have been proposed to fix it. Unfortunately, 3D-Secure is found to be inconvenient and easy to break~\cite{DBLP:conf/fc/MurdochA10}. Tokenization framework offers a great alternative by replacing original card information with temporary tokens during a transaction. However, card information can still be stolen during account setup phase at a poorly secured merchant.
Other unregulated digital financial services are also reported to be insecure~\cite{DBLP:conf/uss/ReavesSBTB15}. In~\cite{DBLP:conf/uss/ReavesSBTB15}, the authors showed that \textit{branchless banking} apps that leverage cellular networks to send/receive cashes are also vulnerable due to flaws such as skipping SSL/TLS certificate validation, and using insecure cryptographic primitives. 
\section{Conclusion}
Our study shows that the PCI data security standard (PCI DSS) is comprehensive, but there is a big gap between the specifications and their real-world enforcement. Our testbed experiments revealed that the vulnerability screening capabilities of some approved scanning vendors (ASV) are inadequate.  5 of the 6 PCI scanners are not compliant with the ASV scanning guidelines. All 6 PCI scanners would certify e-commerce websites that remain vulnerable. Our measurement on 1,203 e-commerce websites shows that 86\% of the websites have at least one type of vulnerability that should disqualify them as non-compliant. Our future work is to a design minimum-footprint black-box scanning method.

\section{Acknowledgment}

This project was supported in part by NSF grants CNS-1717028, CNS-1750101 and OAC-1541105, ONR Grant ONR-N00014-17-1-2498.


\bibliographystyle{acm}
\bibliography{paper}

\section*{Appendix}

  \begin{figure}[!ht]
  	\centering
  	\includegraphics[width=0.50\textwidth]{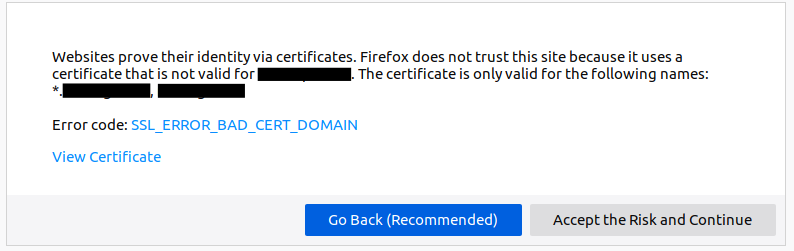}
  	\vspace{-10pt}
  	\caption{An example of wrong hostname in the certificate. The domain (\textit{a*****.***}) uses a certificate that is issued for a different domain name (\textit{*.n*****.***}).}\label{fig:wronghostname}
 \end{figure}
 
 \begin{figure}[!ht]
  	\centering
  	\includegraphics[width=0.50\textwidth]{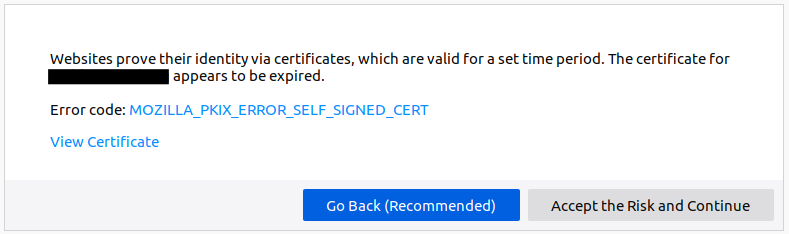}
  	\vspace{-10pt}
  	\caption{Self-signed certificate used by (\textit{r*****.***}), a website that accepts payment cards for donations.}\label{fig:selfsigned}
 \end{figure}
 
  \begin{figure}[!ht]
  	\centering
  	\includegraphics[width=0.50\textwidth]{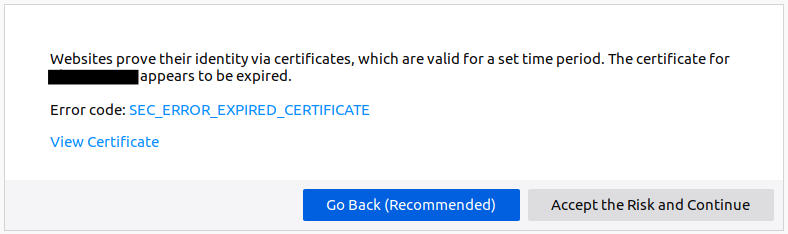}
  	\vspace{-10pt}
  	\caption{(\textit{u*****.***}) uses expired certificates by default and redirects users to a secure sub-domain with proper certificate during payment.}\label{fig:expired}
 \end{figure}

  \begin{figure}[!ht]
  	\centering
  	\includegraphics[width=0.40\textwidth]{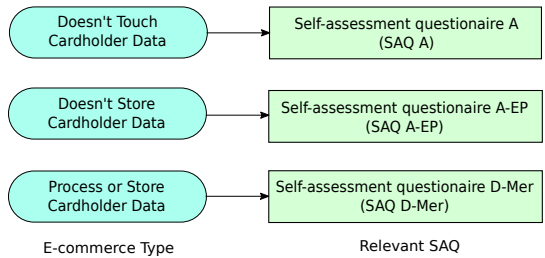}
  	\caption{Self-Assessment Questionnaires (SAQs) for different types of e-commerce merchants.}\label{fig:saq_chart1}
 \end{figure}

\begin{figure*}[hbt]
 
\scriptsize
\begin{tabular}{@{}|>{\centering}p{8cm}|c|c|c|c|c|@{}}
\hline
\multirow{5}{*}{\normalsize \begin{tabular}[c]{@{}l@{}}The card verification code or value (three digit or four-\\ digit number printed on the front or back of a payment \\ card) is not stored after authorization?\end{tabular}} & \cellcolor{green!25} & \cellcolor{green!25} & \cellcolor{green!25} & \cellcolor{green!25} & \cellcolor{green!25} \\ 
&  \cellcolor{green!25}   &  \cellcolor{green!25}            &  \cellcolor{green!25}  &  \cellcolor{green!25}   &  \cellcolor{green!25}          \\ 
& \multirow{-1}{*}{\begin{tabular}[c]{@{}c@{}} \cellcolor{green!25}Yes\end{tabular}}    &  \multirow{-3}{*}{\begin{tabular}[c]{@{}c@{}}Yes \\ with \\ \cellcolor{green!25} CCW\end{tabular}} & \multirow{-1}{*}{\begin{tabular}[c]{@{}c@{}}\cellcolor{green!25}No\end{tabular}}   &  \multirow{-1}{*}{\begin{tabular}[c]{@{}c@{}}\cellcolor{green!25}N/A\end{tabular}}   &  \multirow{-2}{*}{\begin{tabular}[c]{@{}c@{}}Not \\ \cellcolor{green!25} Tested\end{tabular}}      \\ \cline{2-6}
&  \multirow{2}{*}{\begin{tabular}[c]{@{}l@{}}\Square\end{tabular}}   &    \multirow{2}{*}{\begin{tabular}[c]{@{}l@{}}\Square\end{tabular}}  & \multirow{2}{*}{\begin{tabular}[c]{@{}l@{}}\Square\end{tabular}} & \multirow{2}{*}{\begin{tabular}[c]{@{}l@{}}\Square\end{tabular}} & \multirow{2}{*}{\begin{tabular}[c]{@{}l@{}}\Square\end{tabular}} \\
&     &              &    &     &            \\ \hline
\end{tabular}
\caption{A sample question from the Self-Assessment Questionnaire D (SAQ D)~\cite{DBLP:books/saq/saqd}. ``Yes with CCW'' means ``the expected testing has been performed, the requirement has been met with the assistance of a compensating control, and a Compensating Control Worksheet (CCW) is required to be submitted along with the questionnaire'' .}\label{fig:saqd}
\end{figure*}

\subsection*{Implementation Details of \checker{}}\label{sec:pci-checker-impl}

\checker{} follows a series of rules for vulnerability testing. The index of the rules matches with the testing cases discussed in the paper. As described in the paper, we only focus on a subset of test cases that do not disrupt or cause any negative impact to the remote servers (for ethical considerations). The implementation details are as follows.  

\smallskip

\noindent
{\em Rule 2. Database port detection.} For database port detection, we choose to probe for Mysql port\footnote{We do not probe for multiple ports to avoid suspicions for possible port scanning. However, a similar technique can be used to probe for other databases.}. The reason for choosing Mysql port are {\em i)} Mysql is among the top three (Mysql, Oracle, Microsoft SQL Server) most popular databases in the world~\cite{Mysql:popularity}; {\em ii)} Mysql is free; and {\em iii)} it supports a wide range of programming languages. The access to Mysql port ({\em e.g.}, 3306) is disabled by default. It is very dangerous to enable remote access to Mysql database for an arbitrary client. We check the Mysql port using \textit{nc}~\cite{netcattool}, which is a Unix utility tool that reads and writes data across network connections using the TCP or UDP protocol.

\smallskip
\noindent
{\em Rule 5. Default Mysql user/password detection.} If the Mysql database of a website is remotely accessible, we further check for the default username and password. A typical Mysql installation has a user ``\textit{root}'' with an empty password, unless it is otherwise customized or disabled. As such, we run a Mysql client to connect to the remote host using the default username and password. \checker{} terminates the connection immediately and raises an alert if the attempt is successful.

\smallskip
\noindent
{\em Rules 3 \& 19. Checking OpenSSH's availability and version.} We use \textit{nc}~\cite{netcattool} to connect with port 22 of the remote OpenSSH server. If OpenSSH runs on port 22, then it will return the server information ({\em e.g.}, OpenSSH version, OS type, OS version). We parse the returned information to determine the version of the OpenSSH server. We consider any installation versions before \textit{OpenSSH\_7.6} as vulnerable.

\smallskip
\noindent
{\em Rules 29 \& 33. Checking HTTP header information.} Extracting HTTP information does not require the rich browser functionality. We use Java net URL APIs to open HTTP connections for extracting HTTP headers. For case 29, we raise a warning only if we detect that the ``\texttt{Server}'' header contains server name and version. For case 33, we raise a warning if any of the four security header ({\em i.e.}, \texttt{X-Frame-Options}, \texttt{X-XSS-Protection}, \texttt{Strict-Transport-Security}, \texttt{X-Content-Type-Options}) is missing. 

\smallskip
\noindent
{\em Rule 7. Sensitive information over HTTP.} We tested whether all the HTTP traffic is redirected to HTTPS by default. We open an HTTP connection with the server and follow the redirection chain. If the server doesn't redirect to HTTPS, we raise an alert. We use Java net URL APIs to implement this test case.

\smallskip
\noindent
{\em  Rules 18 \& 13. TLSv1.0 and weak cipher negotiation.} We use OpenSSL's \textit{s\_client} tool to establish a SSL/TLS connection using TLSv1.0 protocol. \checker{} raises a warning if the connection is successful. We also use \textit{s\_client} to negotiate the ciphersuite with the remote server. \checker{} raises a warning if we successfully negotiate with a ciphersuite that contains a weak cipher ({\em i.e.}, IDEA, DES, MD5).

\smallskip
\noindent
{\em Rules 12, 14, 15, 16 \& 17. Retrieving and examining the certificate.} We use OpenSSL's \textit{s\_client} tool to retrieve the SSL certificate of a remote server. To parse the certificate, we use APIs from \textit{java.security.cert} package. To check whether a certificate is self-signed (Case 12), we used the public key of the certificate to verify the certificate itself. To check whether the certificate is expired, we use the \textit{checkValidity()} method of \textit{X509Certificate} API (Case 14). If the subject domainname (DN) or any alternate DN of a certificate doesn't match with the server domainname, then \checker{} raises an alert (Case 15). Regarding the public key sizes for factoring modulus ({\em e.g.}, RSA, DSA), the discrete logarithm ({\em e.g.}, Diffie-Hellman), and the elliptic curve ({\em e.g.}, ECDSA) based algorithms, NIST recommends them to be 2048, 224 and 224 bits, respectively~\cite{NIST:keylength}. \checker{} raises alert if the key size is smaller than what is recommended (Case 16).  If the signing algorithm uses any of the weak hashing algorithms ({\em e.g.}, MD5, SHA, SHA1, SHA-1), \checker{} raises warnings (Case 17).

\smallskip
\noindent
{\em Rule 25. Script source integrity check.} A website is expected to check the integrity of any JavaScript code that is loaded externally to the browser. To enable script source integrity check, a server can use the ``\textit{integrity}'' attribute of the \textit{script} tag. In the ``\textit{integrity}'' attribute, the server should mention the hashing algorithm and the hash value of the script that should be used to check the integrity. \checker{} downloads the index page of a website. After that, it collects all the \textit{script} tags, and checks if the \textit{script} tags contain any external URL (excluding the website's CDN URLs). Then it looks for the \textit{integrity} attribute for the scripts loaded from external URLs, and raises alert if the \textit{integrity} attribute is missing. We only perform this test for the index page (instead of all the pages) of a website to keep the test lightweight. The number of vulnerable websites detected by this test can only be interpreted as a lower bound.


\begin{table*}[!h]
\begin{center}
\caption{A summary of the guidelines for ASV scanners~\cite{DBLP:books/asv/guideline}. In the fourth column, we show the categories that are required to be fixed. ``$^*$" means that in the SSL/TLS category, all the vulnerabilities are required to be fixed, except case 18.} \label{asv:scanning:guideline}
\footnotesize
\begin{tabular}{@{}llll@{}}
\toprule
Target Component & Expectation & Test-cases & Must fix?\\ \midrule

Firewalls and Routers     & \begin{tabular}[c]{@{}l@{}}1. Must scan all network devices such as firewalls and external routers.\\ 2. Must test for known vulnerabilities and patches.\end{tabular} & 1  & Yes\\ \hline

Operating Systems         & \begin{tabular}[c]{@{}l@{}}1. Must scan to determine the OS type and version.\\ 2. An unsupported OS must be marked as an automatic failure.\end{tabular} & - & Yes \\ \hline

Database Servers          & \begin{tabular}[c]{@{}l@{}}1. Must test for open access to databases from the Internet.\\ 2. If found - must be marked as an automatic failure (Req. 1.3.6)\end{tabular} & 2 & Yes\\ \hline

Web Servers   & \begin{tabular}[c]{@{}l@{}}1. Must be able to test for all known vulnerabilities and configuration issues.\\ 2. Report if directory browsing is observed.\end{tabular} &  30 & Yes\\ \hline

Application Servers       & 1. Must be able to test for all known vulnerabilities and configuration issues. &  29, 33 & Yes\\ \hline

Common Web Scripts  & 1. Must be able to find common web scripts (e.g., CGI, e-commerce, etc.). & - & Yes \\ \hline

Built-in Accounts         & \begin{tabular}[c]{@{}l@{}}1. Look for default username/passwords in routers, firewalls, OS and web or DB servers.\\ 2. Such vulnerability must be marked as an automatic failure. (Req 2.1)\end{tabular} & 5, 6 & Yes\\ \hline

DNS and Mail Servers      & \begin{tabular}[c]{@{}l@{}}1. Must be able to detect the presence \\
2. Must test for known vulnerabilities and configuration issues\\ 3. Report if a vulnerability is observed (automatic failure for DNS server vulnerabilities).\end{tabular} & - & Yes\\ \hline

Virtualization components & 1. Must be able to test for all known vulnerabilities & - & Yes \\  \hline

Web Applications          & \begin{tabular}[c]{@{}l@{}}Must find common vulnerabilities (automatically/manually) including the following:\\ 1. Unvalidated parameters that might lead to SQL injection.\\ 2. Cross-site scripting (XSS) flaws\\ 3. Directory traversal vulnerabilities\\ 4. HTTP response splitting/header injection\\ 5. Information leakage: phpinfo(), Insecure HTTP methods, detailed error msg\\ 6. If found any of the above must be marked as an automatic failure\end{tabular} & \begin{tabular}[c]{@{}l@{}}21, 22, 23, \\ 24, 25, 26, 27, \\ 28, 31, 32 \end{tabular}  & Yes \\ \hline

Other Applications        & 1. Must test for known vulnerabilities and configuration issues & 20 & Yes \\ \hline

Common Services        & 1. Must test for known vulnerabilities and configuration issues & 19 & Yes \\ \hline

Wireless Access Points    & \begin{tabular}[c]{@{}l@{}}1. Must be able to detect wireless access points\\ 2. Must test and report known vulnerabilities and configuration issues\end{tabular} & - & Yes \\ \hline

Backdoors/Malware       & \begin{tabular}[c]{@{}l@{}}1. Must test for remotely detectable backdoors/malware \\ 2. Report automatic failure if found one \end{tabular} & - & Yes \\ \hline

SSL/TLS & \begin{tabular}[c]{@{}l@{}}Must find:\\ 1. Various version of crypto protocols \\ 2. Detect the encryption algorithms and encryption key strengths \\
3. Detect signing algorithms used for all server certificates \\
4. Detect and report on certificate validity \\
5. Detect and report on whether CN matches the hostname \\
6. Mark as failure if supports SSL or early versions of TLS. \end{tabular} & 12-18 & Yes$^*$ \\ \hline

\begin{tabular}[c]{@{}l@{}}Anonymous Key agreement \\
Protocol \end{tabular} & \begin{tabular}[c]{@{}l@{}}1. Must identify protocols allowing anonymous/non-authenticated cipher suites\\ 2. Report if found one \end{tabular} & -  & Yes \\ \hline

Remote Access     & \begin{tabular}[c]{@{}l@{}}1. Must be able to detect remote access software \\ 2. Must report if one is detected. \\
3. Must test and report known vulnerabilities and configuration issues \end{tabular} & \begin{tabular}[c]{@{}l@{}}3, 4 \\ 19, 20\end{tabular} & Yes \\ \hline

Point-of-sale (POS) Software & \begin{tabular}[c]{@{}l@{}}1. Should look for POS software\\ 2. If found - ask for justification \end{tabular} & - & No \\ \hline

\begin{tabular}[c]{@{}l@{}}Embedded links or code \\
from out-of-scope domains \end{tabular} & \begin{tabular}[c]{@{}l@{}}1. Should look for out-of-scope links/code \\ 2. If found - ask for justification \end{tabular} & - & No \\ \hline

\begin{tabular}[c]{@{}l@{}}Insecure Services/ \\
industry-deprecated protocols \end{tabular} & \begin{tabular}[c]{@{}l@{}} 1. If found one - ask for justification \end{tabular} & - & No \\ \hline

Unknown services & \begin{tabular}[c]{@{}l@{}}1. Should look for unknown services and report if found \end{tabular} & - & No \\\bottomrule
\end{tabular}
\end{center}
\end{table*}

\smallskip
\noindent
{\em Rule 30. Checking for browsable directories.} We check whether the directories are browsable in a website. To avoid redundant traffic, we reuse the collected JavaScript script URLs for case 25. We then examine the common parent directory of all the internal URLs. Finally, we send a GET request to fetch the content of the directory. If directory browsing is enabled, the server will return a response with code 200 with a page containing the listing of files and directories of the specified path. Otherwise, it should return an error response code ({\em e.g.}, 404 - not found, 403 - Forbidden). This test only determines if a directory is browsable. We never store any of the returned pages during the test. 

\begin{table*}[!hbt]
\centering
\caption{Specifications defined by the PCI Security Standard Council (SSC) along with their targets, evaluators, assessors and whether it is enforced by SSC. ``COTS" stands for Commercial Off-The-Shelf.}~\label{pcissc:specs}
\footnotesize 
\begin{tabular}{@{}lllll@{}}
\toprule
PCI Specifications    & Target(s) & Evaluator(s) & Assessor(s): Type & Required? \\ \midrule

Data Security Standard (DSS)~\cite{DBLP:books/pcidss/req}  & \begin{tabular}[c]{@{}l@{}}Merchant, Acquirer Bank, Issuer Bank,\\ Token Service Provider,\\ Service Provider\end{tabular}                                      & \begin{tabular}[c]{@{}l@{}}Acquirer,\\ Payment Brand\end{tabular} & \begin{tabular}[c]{@{}l@{}}QSA: Manual\\ ASV: Automated\end{tabular} & Yes \\ \hline

Card Production and Provisioning (CPP)~\cite{DBLP:books/cpp/logical, DBLP:books/cpp/physical}  & \begin{tabular}[c]{@{}l@{}}Card Issuer, \\ Card Manufacturer, \\ Token Service Provider \\\end{tabular} & Payment Brand & CPP-QSA: Manual  & Yes \\ \hline

Payment Application DSS (PA DSS)~\cite{DBLP:books/pa/dss}   & PA Vendors  & PA-QSA & PA-QSA: Manual & Optional  \\ \hline

Point-to-Point Encryption (P2PE)~\cite{DBLP:books/pci/p2poe} & POS Device Vendors & P2PE-QSA  & P2PE-QSA: Manual  & Optional  \\ \hline

PIN Transaction Security (PTS)~\cite{DBLP:books/pts/poi, DBLP:books/pts/hsm}  & PIN Pad Vendors  & PTS Labs & PTS Labs: Manual  & Optional  \\ \hline

3-D Secure (3DS)~\cite{DBLP:books/pci/3ds}  & \begin{tabular}[c]{@{}l@{}}3DS Server,\\ 3DS Directory Server,\\ 3DS Access Control Server\end{tabular}     & Payment Brand  & 3DS-QSA: Manual  &  Optional \\ \hline

Software-Based PIN Entry on COTS (SPoC)~\cite{DBLP:books/pci/spoc} & PIN-based Cardholder verification method (CVM) Apps & SPoC Labs  & SPoC Labs: Manual  & Optional  \\ \hline
Token Service Provider (TSP)~\cite{DBLP:books/pci/tsp}  & Token Service Providers  & P2PE-QSA  & P2PE-QSA: Manual & Optional  \\ \bottomrule
\end{tabular}
\end{table*}

\smallskip
\noindent
{\em Rule 31. HTTP TRACE supported.} HTTP TRACE method is used for diagnostic purposes. If it is enabled, the web server will respond to a request by echoing in its response the exact request that it has received. In~\cite{grossman2003cross}, the author has shown that HTTP TRACE can be used to steal sensitive information ({\em e.g.}, cookie, credentials). To examine the HTTP TRACE configuration, we send a HTTP request by setting the method to \texttt{TRACE}. If the TRACE method is enabled by the server, the server will echo the request in the response with a code 200.


\begin{table*}[!h]
\renewcommand{\arraystretch}{1.0}

\caption{PCI DSS requirements are presented with expected testing (from SAQ D-Mer) and the potential test-cases that can be used to evaluate the ASV scanning.}\label{test-cases}
\scriptsize
\begin{tabular}{|l|l|l|l|}
\hline
No. & Requirement  & Expected Testing & Testcase \\ \hline
1.1 & Formalize testing when firewall configurations change    & \begin{tabular}[c]{@{}l@{}}1. Review current network diagram\\ 2. Examine network configuration\end{tabular}                                                         & N/A \\ \hline

1.2 & \begin{tabular}[c]{@{}l@{}}Build a firewall to restrict "untrusted" traffic\\ to cardholder data environment\end{tabular}                                     & \begin{tabular}[c]{@{}l@{}} 1. Review firewall and router config\\ 2. Examine firewall and router config\end{tabular}  & \begin{tabular}[c]{@{}l@{}} 1. Enable/disable firewall.\end{tabular} \\ \hline

1.3 & \begin{tabular}[c]{@{}l@{}}Prohibit direct public access between Internet\\ and cardholder data environment\end{tabular}                                   & 1. Examine firewall and router config & \begin{tabular}[c]{@{}l@{}} 2. Expose Mysql to the Internet \\ 3. SSH over public Internet \\ 4. Remote access to PhpMyadmin \end{tabular}  \\ \hline

1.4 & \begin{tabular}[c]{@{}l@{}}Install a firewall on computers that have connectivity\\  to the Internet and organization's network\end{tabular}   & 1. Examine employee owned-devices & N/A  \\ \hline

2.1 & \begin{tabular}[c]{@{}l@{}}Always change vendor-supplied defaults before\\ installing a System on the network\end{tabular}                                   &  \begin{tabular}[c]{@{}l@{}}1. Examine vendor documentations\\ 2. Observe system configurations\end{tabular}                                                          & \begin{tabular}[c]{@{}l@{}}5. Use default DB user/password\\ 6. Use default Phpmyadmin user/password\end{tabular} \\ \hline

2.2 & \begin{tabular}[c]{@{}l@{}}Develop a configuration standards for all system\\ components that address all known security vulnerabilities.\end{tabular}  & \begin{tabular}[c]{@{}l@{}}1. Examine vendor documentations\\ 2. Observe system configurations\end{tabular}  & N/A \\ \hline

2.3 & \begin{tabular}[c]{@{}l@{}} Encrypt using Strong cryptography all non-console\\ administrative access such as browser/web-based\\ management tools\end{tabular} & \begin{tabular}[c]{@{}l@{}}1. Examine system components\\ 2. Examine system configurations\\ 3. Observe an administrator log on\end{tabular}                         & \begin{tabular}[c]{@{}l@{}}7. Sensitive information over HTTP\\ \end{tabular}                                            \\ \hline
2.4 & \begin{tabular}[c]{@{}l@{}} Shared hosting providers must also comply \\ with PCI DSS requirements\end{tabular}  & 1. Examine system inventory & N/A\\ \hline

3.1 & Establish cardholder data retention and disposal policies  & 1. Review data retention and disposal policies  & N/A                                                                                                                                                \\ \hline
3.2 & \begin{tabular}[c]{@{}l@{}}Do not store sensitive authentication data\\  (even it is encrypted)\end{tabular} & \begin{tabular}[c]{@{}l@{}}1. Examine system configurations\\ 2. Examine deletion processes\end{tabular}                              & 8. Store CVV in DB \\ \hline

3.3 & Mask PAN when displayed & \begin{tabular}[c]{@{}l@{}}1. Examine system configurations\\ 2. Observe displays of PAN\end{tabular} & 9. Show unmask PAN  \\ \hline

3.4 & Render PAN unreadable anywhere it is stored & \begin{tabular}[c]{@{}l@{}}1. Examine data repositories\\ 2. Examine removable media\\ 3. Examine audit logs\end{tabular} & \begin{tabular}[c]{@{}l@{}}10. Store plain-text PAN (OpenCart)\\ \end{tabular} \\ \hline

3.5 & \begin{tabular}[c]{@{}l@{}}Secure keys that are used to encrypt stored\\ cardholder data or other keys\end{tabular}  & \begin{tabular}[c]{@{}l@{}}1. Examine system configurations\\ 2. Examine key storage locations\end{tabular}  & 11. Use hardcoded key for encrypting PAN \\ \hline

3.6 & Document all key-management process & 1. Review key-management procedures  & N/A \\ \hline

4.1 & \begin{tabular}[c]{@{}l@{}} Use strong cryptography and security protocols\\ during transmission of cardholder data.\end{tabular}                             & 1. Review system configurations   & \begin{tabular}[c]{@{}l@{}}12. Use self-signed certificate\\ 13. Use insecure block cipher \\ 14. Use Expired certificate\\ 15. Use cert. with wrong hostname\\ 16. Use 1024 bit DH modulus. \\ 17. Use weak hash in SSL certificate \\ 18. Use TLSv1.0 \end{tabular} \\ \hline

4.2 & \begin{tabular}[c]{@{}l@{}}Never send PAN over unprotected user\\ messaging technologies.\end{tabular} & 1. Review policies and procedures & N/A  \\ \hline

5.1 & Deploy anti-virus software on all systems & \begin{tabular}[c]{@{}l@{}}1. Examine system configurations\ 2. Interview personnel\end{tabular}  & N/A \\ \hline

5.2 & \begin{tabular}[c]{@{}l@{}} Ensure all anti-virus mechanisms are current,\\ running and generating audit log\end{tabular} & \begin{tabular}[c]{@{}l@{}} 1. Examine anti-virus configurations\\ 2. Review log retention process\\ 3. Examine system configurations\end{tabular}                    & N/A \\ \hline

6.1 & \begin{tabular}[c]{@{}l@{}}Ensure that all system components are protected\\ from known vulnerabilities\end{tabular} & \begin{tabular}[c]{@{}l@{}} 1. Examine system components \\ 2. Compare the list of security patches\end{tabular} & \begin{tabular}[c]{@{}l@{}}19. Use vulnerable of OpenSSH \\ 20. Use vulnerable PhpMyadmin \end{tabular}  \\ \hline

6.2 & \begin{tabular}[c]{@{}l@{}}Establish a process to identify and assign risk\\ to newly discovered security vulnerabilities\end{tabular} & 1. Review policies and procedures  & N/A                                                                                                                                                \\ \hline
6.3 & \begin{tabular}[c]{@{}l@{}} Develop software applications in accordance\\ with PCI DSS and industry best practices\end{tabular}  & 1. Review software development process  & N/A\\ \hline

6.4 & \begin{tabular}[c]{@{}l@{}} Follow change control processes and procedures\\ for all changes to system components\end{tabular}                                 & 1. Review change control process   & N/A \\ \hline

6.5 & \begin{tabular}[c]{@{}l@{}}Develop applications based on secure coding\\ guidelines and review custom application code\end{tabular} & 1. Review software-development policies  & \begin{tabular}[c]{@{}l@{}} 21. Implant SQL injection in admin login\\ 22. Implant SQL injection in customer login \\ 23. Disable password retry limit \\ 24. Disable restriction on password length. \\ 25. Use JS from external source insecurely \\ 26. Do not hide program crashes \\ 27. Implant XSS\\ 28. Implant CSRF \end{tabular}                               \\ \hline
6.6 & \begin{tabular}[c]{@{}l@{}}Ensure all public-facing applications are\\  protected against known attacks\end{tabular} & 1. Examine system configuration  & \begin{tabular}[c]{@{}l@{}} 29. Present server info in security Headers. \\ 30. Browsable web directories. \\ 31. Enable HTTP Trace/Track \\ 32. Enable phpinfo() \\ 33. Disable security headers\end{tabular}  \\ \hline

7   & \begin{tabular}[c]{@{}l@{}}Restrict access to cardholder data \\ based on roles\end{tabular} & \begin{tabular}[c]{@{}l@{}}1. Examine access control policy\\ 2. Review vendor documentation\\ 3. Examine system configuration\\ 4. Interview personnel\end{tabular} & N/A                                                                                                                                                \\ \hline
8.4\footnotemark & \begin{tabular}[c]{@{}l@{}}Render all passwords unreadable during storage\\ and transmission for all system components\end{tabular} & 1. Examine system configuration   & \begin{tabular}[c]{@{}l@{}} 34. Store unsalted customer passwords\\ 35. Store plaintext passwords\end{tabular} \\ \hline

9   & Restrict physical access to cardholder data & \begin{tabular}[c]{@{}l@{}}1. Observe process\\ 2. Review policies and procedures\\ 3. Interview personnel\end{tabular}                                              & N/A                                                                                                                                                \\ \hline
10  & \begin{tabular}[c]{@{}l@{}}Track and monitor all access to network\\ resource and cardholder data\end{tabular} & \begin{tabular}[c]{@{}l@{}}1. Interview personnel\\ 2. Observe audit logs\\ 3. Examine audit log settings\end{tabular}                                               & N/A                                                                                                                                                \\ \hline
11  & Regularly test security systems and processes  & \begin{tabular}[c]{@{}l@{}}1. Interview personnel\\ 2. Examine scope of testing\\ 3. Review results of ASV scans\end{tabular}                                        & N/A                                                                                                                                                \\ \hline
12  & \begin{tabular}[c]{@{}l@{}}Maintain a policy that addresses information \\ security for all personnel\end{tabular}  & \begin{tabular}[c]{@{}l@{}}1. Review formal risk assessment\\ 2. Review security policy\\ 3. Interview personnel.\end{tabular}                                       & N/A                                                      \\ \hline
\end{tabular}
\vspace{5pt}
\linebreak
\footnotesize $^{9}$ Other requirements under 8 are not testable.
\end{table*}

\end{document}